\documentclass[sigconf]{acmart}

\AtBeginDocument{%
  \providecommand\BibTeX{{%
    \normalfont B\kern-0.5em{\scshape i\kern-0.25em b}\kern-0.8em\TeX}}}

\copyrightyear{2025}
\acmYear{2025}
\setcopyright{cc}
\setcctype{by}
\acmConference[CHI '25]{CHI Conference on Human Factors in Computing Systems}{April 26-May 1, 2025}{Yokohama, Japan}
\acmBooktitle{CHI Conference on Human Factors in Computing Systems (CHI '25), April 26-May 1, 2025, Yokohama, Japan}\acmDOI{10.1145/3706598.3714270}
\acmISBN{979-8-4007-1394-1/25/04}

\usepackage{graphicx}
\usepackage{mathtools}
\usepackage{multirow}
\usepackage{enumitem}
\usepackage{subcaption}
\usepackage{colortbl}
\usepackage{xcolor}

\begin{document}

\title[Effects of Increasing Task Complexity on the Effectiveness of Data Storytelling]{"Piecing Data Connections Together Like a Puzzle": Effects of Increasing Task Complexity on the Effectiveness of Data Storytelling Enhanced Visualisations}

\author{Mikaela Elizabeth Milesi}
\authornote{Both authors contributed equally to this research.}
    \orcid{0009-0002-0910-9822}
    \affiliation{%
        \institution{Monash University}
        \city{Melbourne}
        \state{Victoria}
        \country{Australia}
    }
    \email{mikaela.milesi@monash.edu}
    
\author{Paola Mejia-Domenzain}
\authornotemark[1]
    \orcid{0000-0003-1242-3134}
    \affiliation{%
        \institution{EPFL}
        \city{Lausanne}
        \country{Switzerland}
    }
    \email{paola.mejia@epfl.ch}

\author{Laura Brandl}
    \orcid{0000-0001-7974-7892}
    \affiliation{%
        \institution{Department of Psychology, Ludwig-Maximilians-Universität München}
        \city{Munich}
        \country{Germany}
    }
    \email{L.Brandl@psy.lmu.de}

\author{Vanessa Echeverria}
    \orcid{0000-0002-2022-9588}
    \affiliation{%
        \institution{Monash University}
        \city{Melbourne}
        \country{Australia}
    }
    \affiliation{%
        \institution{Escuela Superior Politécnica del Litoral}
        \city{Guayaquil}
        \country{Ecuador}
    }
    \email{vanessa.echeverria@monash.edu}

\author{Yueqiao Jin}
    \orcid{0009-0003-7309-4984}
    \affiliation{%
        \institution{Monash University}
        \city{Melbourne}
        \state{Victoria}
        \country{Australia}
    }
    \email{ariel.jin@monash.edu}

\author{Dragan Gašević}
    \orcid{0000-0001-9265-1908}
    \affiliation{%
        \institution{Monash University}
        \city{Melbourne}
        \state{Victoria}
        \country{Australia}
    }
    \email{dragan.gasevic@monash.edu}

\author{Yi-Shan Tsai}
    \orcid{0000-0001-8967-5327}
    \affiliation{%
        \institution{Monash University}
        \city{Melbourne}
        \state{Victoria}
        \country{Australia}
    }
    \email{yi-shan.tsai@monash.edu}

\author{Tanja Käser}
    \orcid{0000-0003-0672-0415}
    \affiliation{%
        \institution{EPFL}
        \city{Lausanne}
        \country{Switzerland}
    }
    \email{tanja.kaeser@epfl.ch}

\author{Roberto Martinez-Maldonado}
    \orcid{0000-0002-8375-1816}
    \affiliation{%
        \institution{Monash University}
        \city{Melbourne}
        \state{Victoria}
        \country{Australia}
    }
    \email{roberto.martinezmaldonado@monash.edu}


\renewcommand{\shortauthors}{Milesi et al.}

\begin{abstract} 
The emerging concept of \textit{data storytelling} (DS) suggests that enhancing visualisations with annotations and narratives can make complex data more insightful than conventional visualisations. Previous works found that DS-enhanced visualisations are more effective than conventional visualisations for simple tasks like identifying key data points or the main message. However, no previous work has explored the extent to which DS enhancements influence task completion across different levels of cognitive complexity. We address this gap by presenting the results of a study where 128 participants completed tasks based on four visualisations (two line charts and two choropleth maps, either with or without DS elements) spanning a range of complexity based on Bloom's taxonomy, which has been applied in data visualisation to categorise tasks hierarchically from lower to higher-order thinking. 
Results suggest that while DS-enhanced visualisations effectively support lower-order tasks (finding data points and understanding insights), they don't necessarily aid the correct completion of higher-order tasks (\textit{application}, \textit{analysis}, \textit{evaluation} and \textit{creation}). However, DS enhancements improve how efficiently participants complete complex tasks.
\end{abstract}



\maketitle
    \begingroup
        \renewcommand{\thefootnote}{}
        \footnotetext{* These two authors contributed equally to this work.}
        \addtocounter{footnote}{0}
    \endgroup

    \section{Introduction}
Data has become an integral and pervasive component of modern society \cite{dykes2020effective, Zhang2020dataquilt}, influencing sectors ranging from personal decision-making to large-scale governmental policies \cite{martinezmaldonado2020layeredstorytelling, Young-Ho2021, Loerakker2024, hwang2021evidence}. Tasks involving the incorporation of data insights can vary in complexity, from laypersons using data-driven information to learn about new topics \cite{Ferreira2021climate, Zhang2020dataquilt} or about themselves \cite{martinezmaldonado2020layeredstorytelling, Young-Ho2021, Loerakker2024, yen2020,mejia2023dashboards}, to higher-order analysis by data scientists generating insights and visualisations \cite{yang2022, lowe2020requirements}, and governments evaluating data to create policies \cite{hwang2021evidence}. However, the ability to interpret and extract data insights, particularly in tasks of varying cognitive complexity, remains a challenge for many \cite{shao2024datastorytelling, nobre2024, yang2022, ge2024}. As both the availability and complexity of data continue to increase, it is important to support individuals -- particularly those who are not data savvy -- in navigating and interpreting the data \cite{marriott2018immersive, ndukwe2020teaching}.

One of the key roles of data visualisation is to simplify the presentation of data into clear, actionable visuals, enabling people to perform tasks of varying complexity \cite{tufte2001visual}.
Data visualisations are often intended as a communication tool to enable \textit{‘visual discovery’} of insights \cite{ryan2016visual, Wang2020cheat}. Some visualisations are designed and used to enable data \textit{exploration}, where individuals are expected to independently interpret the information and extract insights for analysis \cite{yau2013datapoints, knaflic2015storytelling}. However, this approach relies either on the patterns and insights extracted from the data being readily apparent \cite{yau2013datapoints}, or on individuals possessing the necessary data analysis skills to interpret the information \cite{echeverria2018exploratory}. These are therefore commonly referred to as \textit{exploratory} visualisations, as they are primarily designed for users with data analysis expertise who are exploring unfamiliar datasets to uncover insights \cite{Kosara2013, bravo2020immersive, martinezmaldonado2020layeredstorytelling, iliinsky2011designing}.
In contrast, \textit{explanatory} visualisations serve as an alternative approach whereby judicious design choices and visualisation techniques are applied in advance to enable non-data savvy individuals to make data-driven decisions by directly communicating insights through a data story \cite{ryan2016visual, knaflic2015storytelling}. A technique that enables this explanatory approach is \textit{data storytelling} (DS), whereby information is compressed by adding visual or textual enhancements to conventional visualisations \cite{segel2010narrative, ryan2016visual, fernandeznieto2024editor}.

Theoretical researchers and visualisation experts \cite{segel2010narrative, gershon2001, Daradkeh2021, boy2015storytelling, boy2017showing} have argued that DS helps users more effectively (with greater accuracy) and efficiently (in lesser time) extract insights from data compared to conventional methods \cite{elias2013storytelling, knaflic2015storytelling, zdanovic2022influence, segel2010narrative, gershon2001, krum2013cool, Hicks2009perceptual}. Sensemaking refers to the process of interpreting complex data to uncover patterns, relationships, and insights that might otherwise remain hidden \cite{campos2021makingsense}. DS has been claimed to facilitate this process by using explanatory narratives and visual cues to guide the user \cite{martinezmaldonado2020layeredstorytelling, srinivasan2019augmenting}. Yet, despite the theoretical arguments, empirical findings offer mixed results. Some studies have shown that DS can improve long-term recall \cite{bateman2010}, while others find no significant differences in memory retention compared to conventional visualisations \cite{zdanovic2022influence}. Evidence supporting the \textit{effectiveness} of DS in comprehension tasks comes from \citet{shao2024datastorytelling}, who found that DS visualisations indeed improves insight comprehension, especially for single insights. Yet, findings related to \textit{efficiency} are similarly inconsistent. While small-scale eye-tracking studies \cite{de2014evaluating, Pozdniakov2023how} (with fewer than 30 participants) suggest that DS effectively captures attention, and \citet{echeverria2018exploratory} found that DS-enhanced visualisations reduced interpretation time for educators, a large-scale study \cite{shao2024datastorytelling} showed no improvement in speed. These inconsistencies suggest the need for further research to explore how DS impacts both effectiveness and efficiency in a wide range of tasks that might require different cognitive skills. The primary goal of DS is to draw focus towards salient data points and make key takeaways from data accessible \cite{ryan2016visual}, thus existing studies predominantly focus on lower-order tasks. However, in practice, DS is commonly applied to complex topics that require higher-order thinking, such as politics, economics, or history \cite{seyser2018scrollytelling, morth2023scrollyvis}. This leaves a gap in understanding the impact that DS-enhancements could have on interpreting complex information and completing tasks that require higher-order thinking skills. 

In this study, we aim to examine how DS-enhanced visualisations compare to conventional data visualisations in influencing users' understanding of tasks with varying levels of complexity. To explore this, we conducted a within-subjects controlled experiment where users are presented with tasks spanning the full range of Bloom's taxonomy \cite{krathwohl2002revision}, from low-order thinking skills (\textit{Remember/Identify}, \textit{Understand}) to higher-order tasks (\textit{Apply}, \textit{Analyse}, \textit{Evaluate}, \textit{Create}). Each participant was presented with four visualisations: two line charts and two choropleth maps, both with and without DS elements. An important methodological contribution of this work is the introduction of a novel \textit{Human-in-the-Loop-AI} task generation process. This approach integrates the automation capabilities of large language models with human validation to generate question-based tasks across different levels of Bloom's taxonomy. By systematically aligning cognitive complexity with task design, this process provides a scalable framework for evaluating the cognitive impact of visualisations and can be used by future research to study visualisation efficacy across diverse contexts.
A total of 128 users participated in the study where we addressed the following research questions: To what extent do DS-enhanced visualisations improve \textit{\textbf{effectiveness} (\textbf{RQ1}) and \textit{\textbf{efficiency}} (\textbf{RQ2}) of tasks involving data point identification, understanding, applying, analysing, and evaluating insights compared to conventional visualisations? To what extent do DS-enhanced visualisations improve the \textbf{\textit{quality of a creation task}} output compared to conventional visualisations? (\textbf{RQ3}); and how are visualisations with data storytelling elements \textbf{\textit{perceived}} in comparison to conventional visualisations in terms of preferences, usefulness and utility? (\textbf{RQ4})}

Our findings contribute to the growing body of research on data storytelling by providing empirical evidence on users' performance in a wide range of cognitive tasks. Key insights include:
\begin{itemize}
    \item DS-enhanced visualisations can influence how effectively lower-order tasks, such as finding data points and understanding insights, are completed. However, this influence does not necessarily translate to higher-order tasks where analysis, critical evaluation, or the ability to synthesise information into a novel form are necessary.
    \item For tasks at a higher cognitive level (such as those involving analysis, evaluation, or creation), DS-enhanced visualisations can positively impact the efficiency with which tasks are completed.
    \item Complex writing tasks that were written with DS-enhanced visualisations had less sentences categorised at the \textit{evaluation} level of Bloom's taxonomy. This suggests that individuals may be less inclined to critically engage with the narrative presented by DS-enhanced visualisations.
    \item Participants praised DS-enhancements for providing necessary context and clarity to the data visualisations. However, some individuals still preferred the minimalism and simplicity of conventional visualisations. 
\end{itemize}

    \section{Background}
\subsection{Data Visualisation \& Data Storytelling}
Sensemaking is a key process through which humans interpret and ascribe meaning to information \cite{campos2021makingsense} relying on mental models that serve as functional internal representations of external systems. These mental models preserve structural, behavioural, and data-related properties, enabling reasoning and problem-solving through simulation in working memory \cite{liu2010mental}. \citet{klein2007data} describe sensemaking as a ``\textit{deliberate effort to understand events}'' with key functions including forming explanations, predicting future states, and identifying relationships. Data visualisations, such as line charts or bar charts, support sensemaking by enabling \textit{exploratory} analysis to uncover patterns and insights \cite{Zhang2020dataquilt, knaflic2015storytelling, evergreen2020effective}. These exploratory visualisations are often minimalistic and tailored for audiences familiar with the data context or equipped with analytical skills \cite{martinezmaldonado2020layeredstorytelling, srinivasan2019augmenting}.

Existing research has recognised the potential role that narratives can have on the sensemaking process \cite{figueiras2014narrative, pirolli2005sensemaking}, particularly in aligning external visual representations with users’ mental models \cite{liu2010mental}. To address diverse levels of visualisation literacy and facilitate cognitive offloading \cite{liu2010mental,hullman2011visualization}, \textit{explanatory} visualisations have evolved, incorporating storytelling elements to communicate insights effectively \cite{figueiras2014narrative}. These author-driven “\textit{data stories}” guide audiences through structured narratives that emphasise key findings \cite{segel2010narrative, hullman2011visualization, echeverria2018exploratory}. Data storytelling (DS) combines data, visual elements, and narratives to contextualise and share insights \cite{ryan2016visual, dykes2020effective}. Principles outlined by \citet{echeverria2018exploratory} and \citet{martinezmaldonado2020layeredstorytelling} emphasise guiding audience attention through intentional visualisation, narrative choices, and alignment with specific goals. Complementing this, \citet{hullman2011visualization} argues for the use of rhetorical devices such as framing, annotations, and metonymy guide interpretation, focusing the audience’s attention on key insights to maintain a visual narrative flow \cite{mckenna2017visual}.

DS can take various formats, such as data comics \cite{bach2017emerging, bach2018designpatterns, milesi2024comics}, narrative slideshows \cite{chen2019designing}, data videos \cite{amini2015datavideos, xian2023datavideos}, and annotated charts \cite{echeverria2018exploratory}. Among these, narrative-enhanced visualisations stand out for their effectiveness. Visual features of this format often align with \citet{tufte2001visual}'s “\textit{data-ink ratio},” emphasising components that directly support insight delivery \cite{knaflic2015storytelling, zdanovic2022influence}. Design features proposed by \citet{knaflic2015storytelling} include: 
\begin{enumerate}
    \item \textit{\textbf{Highlighting important data points}} by using visual attributes like colour, shading, or shapes to direct attention to critical insights \cite{ryan2016visual,hullman2011visualization}.
    \item \textit{\textbf{Explanatory titles}} that guide the interpretation of the visualisation by directly communicating the main insight intended by the creator \cite{borkin2016beyond, echeverria2018exploratory}. 
    \item \textit{\textbf{Annotations}} that provide context or explain key takeaways, focusing viewers on significant aspects \cite{hullman2011visualization, knaflic2015storytelling}.
\end{enumerate}

Additionally, DS-enhanced visualisations often employ \textit{\textbf{decluttering}} to simplify the communication of potentially complex ideas by removing non-essential elements, thereby minimising cognitive load and clarifying the narrative \cite{hullman2011visualization, knaflic2015storytelling}.

\subsection{Data Storytelling Effectiveness and Efficiency}
Theoretical researchers and visualisation experts, such as \citet{segel2010narrative}, \citet{gershon2001}, and \citet{Daradkeh2021} have argued that combining narratives with graphics makes visualisations more intuitive and should, in theory, improve comprehension and decision-making. An explanation for this is that DS leverages pre-attentive processing, using visual attributes like colour and shape to guide attention and facilitate quicker understanding of key data points \cite{krum2013cool,Hicks2009perceptual, knaflic2015storytelling}. However, empirical researchers have found mixed evidence regarding the effectiveness and effectiveness of DS-enhanced visualisations.  

Regarding the effectiveness in low cognitive tasks like remembering, \citet{bateman2010} found that visual embellishments, such as colour and textual pointers, did not impact short-term recall compared to minimalist visualisations but did significantly enhance long-term recall. In contrast, \citet{zdanovic2022influence} found no significant differences in memory retention between DS-enhanced and conventional visualisations in either short-term or long-term contexts. Further evidence supporting the effectiveness of DS comes from \citet{shao2024datastorytelling}, who conducted a study to explore how DS-enhancements affect information retrieval and insight comprehension. Their findings suggest that DS visualisations improve the effectiveness of comprehension tasks, particularly those involving single insights, compared to conventional visualisations. Moreover, \citet{stokes2023striking} found that charts with heavy textual annotations were preferred by users and led to better comprehension, particularly when the annotations highlighted statistical or relational components. Similarly, \citet{kong2018} found that framing the titles of visualisations influenced users' perceptions and recall, indicating that even subtle DS elements like title framing can affect how insights are derived. It is noteworthy that these studies were limited to tasks requiring low levels of cognition like finding specific data points or understanding the main take away message of the visualisation. 

The efficiency of DS-enhanced visualisations — how quickly individuals can extract insights — has also been the subject of investigation to some extent. Small-scale eye-tracking studies \cite{de2014evaluating,Pozdniakov2023how} suggest that DS elements attract users' attention more effectively, yet they do not provide direct evidence of whether such elements facilitate quicker comprehension. In an educational context, \citet{echeverria2018exploratory} found that DS elements can reduce the time required to process information, where teachers were able to interpret data faster with DS-enhanced visualisations compared to conventional ones. However, the only large-scale empirical study on efficiency comparing conventional visualisations against those enhanced with DS elements \cite{shao2024datastorytelling} found that while DS elements improved the effectiveness of insight comprehension, they did not necessarily make the process faster.

\subsection{Data Storytelling for Supporting Tasks of Varying Complexity}
The relationship between task complexity and the potential impact that it can have on the effectiveness of visualisations has already been explored in the literature. Many of these prior studies have used Bloom's taxonomy, either partially or fully, as a framework to understand this effect. Bloom's taxonomy is a hierarchical framework that emerged from educational literature \cite{bloom1956} and it has recently been adapted and adopted in information visualisation literature \cite{burns2020evaluate, leerobbins2022learning}. It organises tasks by cognitive complexity, ranging from basic levels of thinking, such as identifying and understanding, to higher-order cognitive processes including analysing, evaluating, and creating. In the revised version of Bloom's taxonomy \cite{krathwohl2002revision}, the category previously referred to as \textit{Knowledge} was renamed \textit{Remember} to reflect the focus on retrieving relevant knowledge. In the context of information visualisation (InfoVis), however, the first level is often referred to as \textit{Identify}, as it better captures the process of recognising and selecting information from visual data \cite{shao2024datastorytelling}. Accordingly, throughout this paper, we will use the term \textit{Identify} to refer to the first level of Bloom's taxonomy.

The first level of Bloom's taxonomy was the key focus of the \citet{zdanovic2022influence} 
exploration into the effects that DS has on the ability to recall information delivered by visualisations. They found that there were no significant differences in recall between conventional and DS-enhanced visualisations.

In their investigation into the effect of DS on information retrieval and insight comprehension, \citet{shao2024datastorytelling} found that there were no significant differences in how efficiently participants were able to successfully answer multiple-choice questions based on DS-enhanced visualisations compared to conventional visualisations. Conversely, they found that the addition of storytelling elements can significantly influence how effectively individuals comprehend data insights, as measured by the number of correct responses. However, this study was limited to the first two levels of Bloom's taxonomy: \textit{Remember/Identify data points} and \textit{Understand}.

In contrast to the aforementioned studies that focus primarily on the lower levels of Bloom's taxonomy, \citet{burns2020evaluate} provided a framework adapted from the full extent of the original taxonomy to evaluate the levels to which participants understood visualisations. Although the visualisations do not explicitly incorporate the design features of DS from \citet{knaflic2015storytelling}, \citet{burns2020evaluate} highlights that the design decisions made when creating visualisations can influence a reader's ability to correctly interpret information. This motivates the exploration of the effectiveness and efficiency of visualisation with DS elements, not only at lower levels of cognitive complexity but also to understand the extent to which the design enhancements impact higher-level tasks.

Recently, the revised Bloom's taxonomy from \citet{krathwohl2002revision} has been used by \citet{adar2021communicative} and \citet{leerobbins2022learning} as part of their research into an explanatory approach called communicative visualisations. Specifically, Bloom's taxonomy inspired the structure of learning objectives, a method of framing the intention of the visualisation in order to guide the audience, as well as evaluate the "success" of the visualisation's design \cite{adar2021communicative}. However, while the effectiveness of the visualisation design is discussed using metrics such as memorability, readability, and user preference, there is minimal discussion on the relationship between the complexity of the task (as measured by Bloom's taxonomy) and the design of the visualisations. 

    \section{Method}
To address the research questions, this paper presents a study comparing pairs of visualisations: conventional versus DS-enhanced, across a range of topics. Tasks were developed based on Bloom's taxonomy to assess different levels of cognitive complexity. A within-subjects design was used, where participants completed both multiple-choice and open-ended question-based tasks for each visualisation pair.
The rest of this section outlines the key elements of the study, including: i) the dataset and materials used; ii) the creation and transformation process for conventional and DS-enhanced visualisations; iii) the development of question-based tasks aligned with Bloom's taxonomy to evaluate the visualisations; iv) the experimental design and procedure; v) participant recruitment and demographics; and vi) the analysis performed for each research question.

\subsection{Dataset and Materials}
\label{dataset}
In order to compare conventional visualisations with those containing DS-enhancements (RQs 1-4), we first needed to identify an appropriate dataset that could be used to extract a set of data stories to communicate.

For this study, we aimed to find a dataset that would be unfamiliar enough to the general public that they would require the visualisations to complete tasks of different complexity, but not so inaccessible that they would be unable to answer questions at all. To this end, we selected open-source data from the \textit{Our World in Data} website. The data stories presented in this study were based on the interconnected topics of \textit{Colonisation}, \textit{Territory Control}, \textit{State Capacity}, and \textit{Tax as a percentage of GDP}, as explained by \citet{herre2023statecap}\footnote{Dataset: \url{https://ourworldindata.org/state-capacity}}. These topics were chosen over alternative datasets because the \textit{Our World in Data} page already contained well-documented insights, reducing the risk of introducing bias or fabricating narratives. Moreover, the topics offered a diverse set of visual representations, including both maps and line charts, alongside detailed endnotes, which provided a rich context for DS-enhancements.

The visualisations in this study (to be fully described below, in Section \ref{design_process}) were created using Python's \textit{Matplotlib} and \textit{Geopandas} library. Additional visual embellishments, such as annotations, were added using the graphic design platform \textit{Canva} \footnote{Canva: \url{https://www.canva.com/}}.

\subsection{Visualisation Design and Transformation Process}
\label{design_process}
\begin{figure*}[h]
    \centering
    \includegraphics[width=1\linewidth]{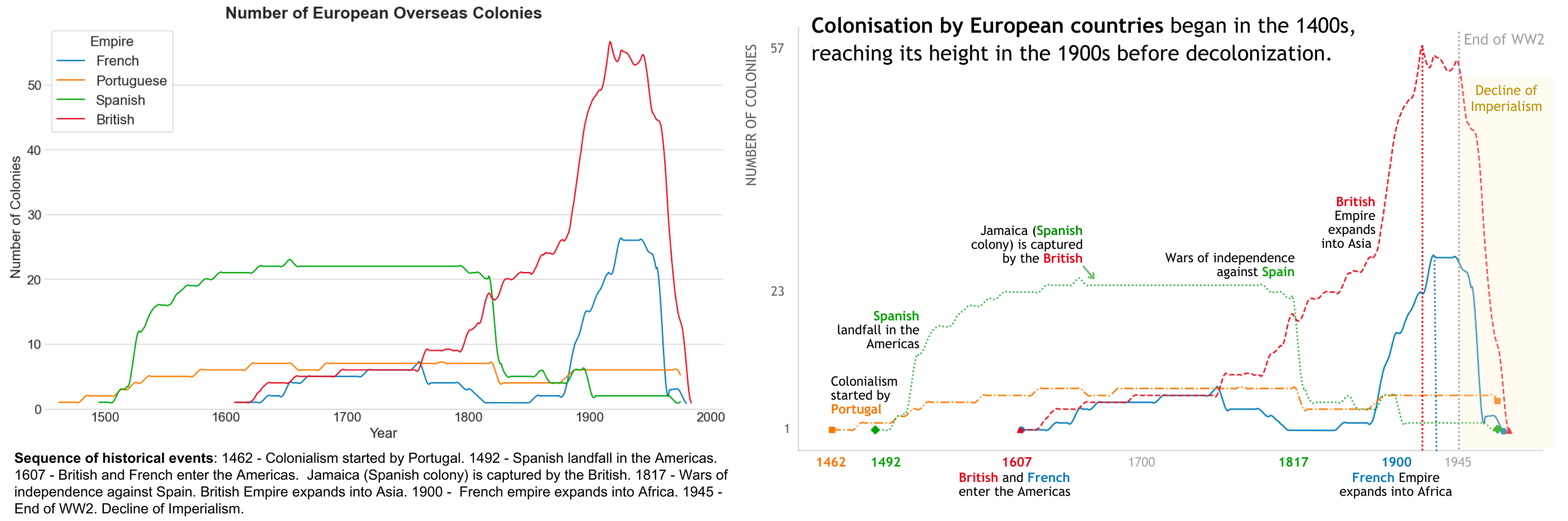}
    \caption{Visualisation \texttt{A} based on the topic of Colonialism. The pair consists of the conventional visualisation \texttt{A-CV} [left] and the DS-enhanced visualisation \texttt{A-DS} [right].}
    \Description{A pair of line charts each depicting the number of European colonies from 1400 to 2000. For both visualisations the number of colonies is the y-axis and time is on the x-axis. The left visualisation is a conventional visualisation. It has a descriptive title, a legend, gridlines, axis labels, and a footnote paragraph detailing key events. The right visualisation is enhanced by data storytelling. It has an explanatory title, no gridlines, minimal text on the axes, and annotations next to key events.}
    \label{fig:visA}
\end{figure*}

\begin{figure*}[htbp]
    \centering
    \includegraphics[width=1\linewidth]{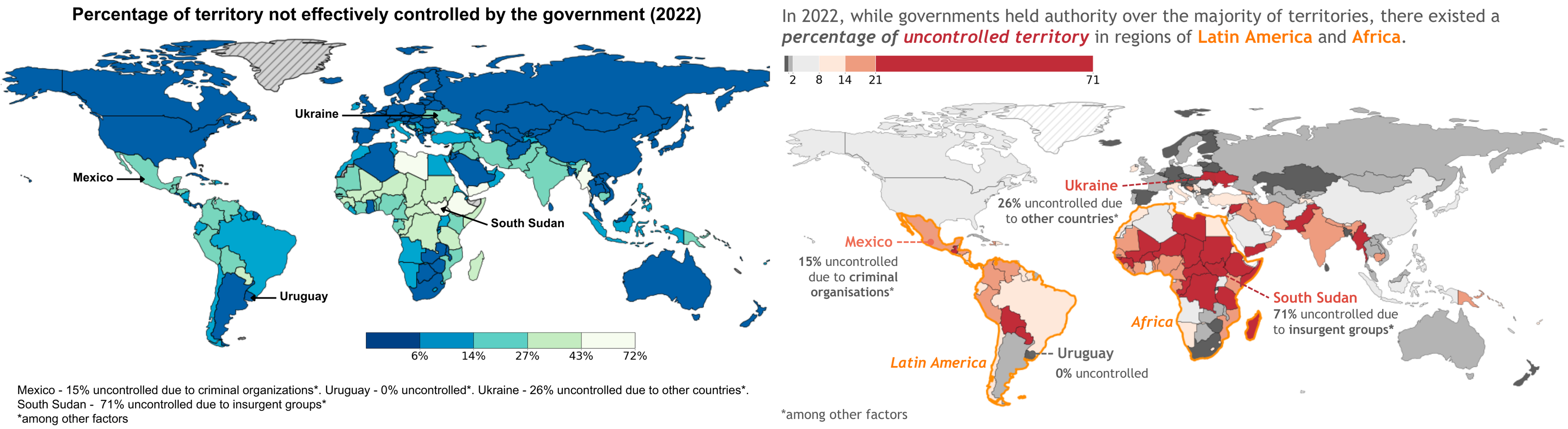}
    \caption{Visualisation \texttt{B} based on the topic of Territory Control. The pair consists of the conventional visualisation \texttt{B-CV} [left] and the DS-enhanced visualisation \texttt{B-DS} [right].}
    \Description{A pair of maps each depicting the percentage of territory not controlled by the government in 2022. The left visualisation is a traditional visualisation with a descriptive title, a legend, labels for particular country names, and a footnote paragraph highlighting key countries with reasons for their percentage of controlled territory. The right visualisation is enhanced by data storytelling. It has an explanatory title and annotations labelling key countries accompanied with a reason for their percentage of controlled territory.}
    \label{fig:visB}
\end{figure*}

\begin{figure*}[htbp]
    \centering
    \includegraphics[width=1\linewidth]{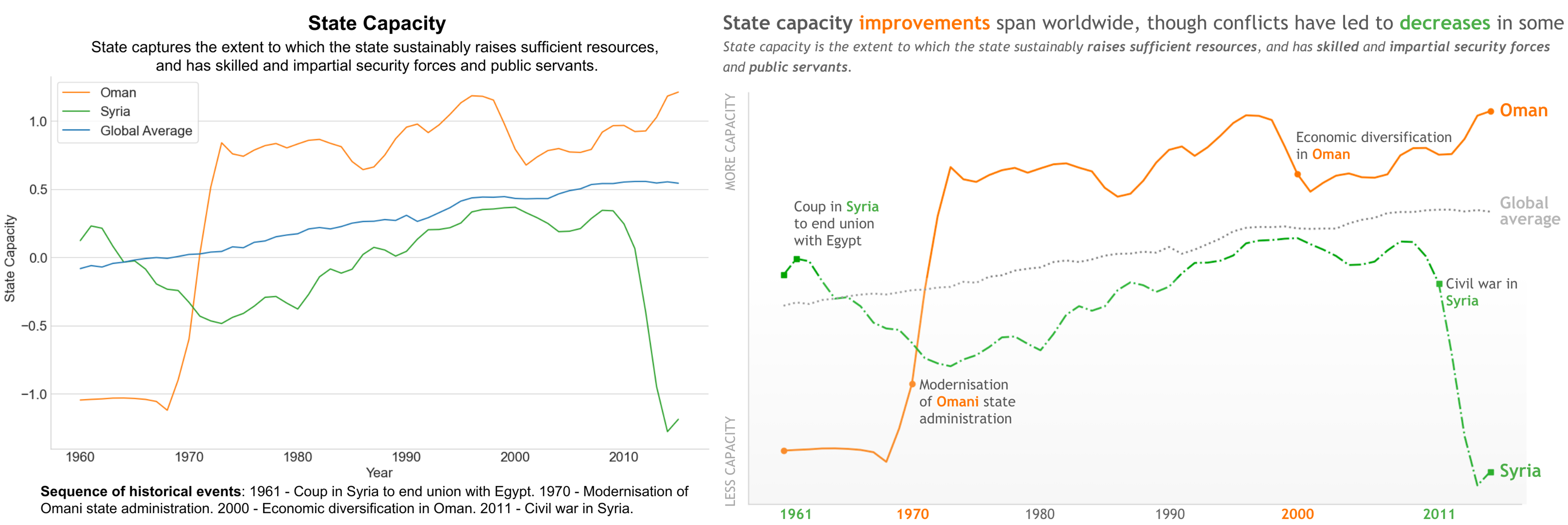}
    \caption{Visualisation \texttt{C} based on the topic of State Capacity. The pair consists of the conventional visualisation \texttt{C-CV} [left] and the DS-enhanced visualisation \texttt{C-DS} [right].}
    \Description{A pair of line charts each depicting the state capacity of Oman, Syria, and the Global Average from 1960 to 2012. For both visualisations state capacity is the y-axis and the year is on the x-axis. The left visualisation is a traditional visualisation. It has a descriptive title, a legend, gridlines, axis labels, and a footnote paragraph detailing key events. The right visualisation is enhanced by data storytelling. It has an explanatory title, no gridlines, minimal text on the axes, and annotations next to key events.}
    \label{fig:visC}
\end{figure*}

\begin{figure*}[htbp]
    \centering
    \includegraphics[width=1\linewidth]{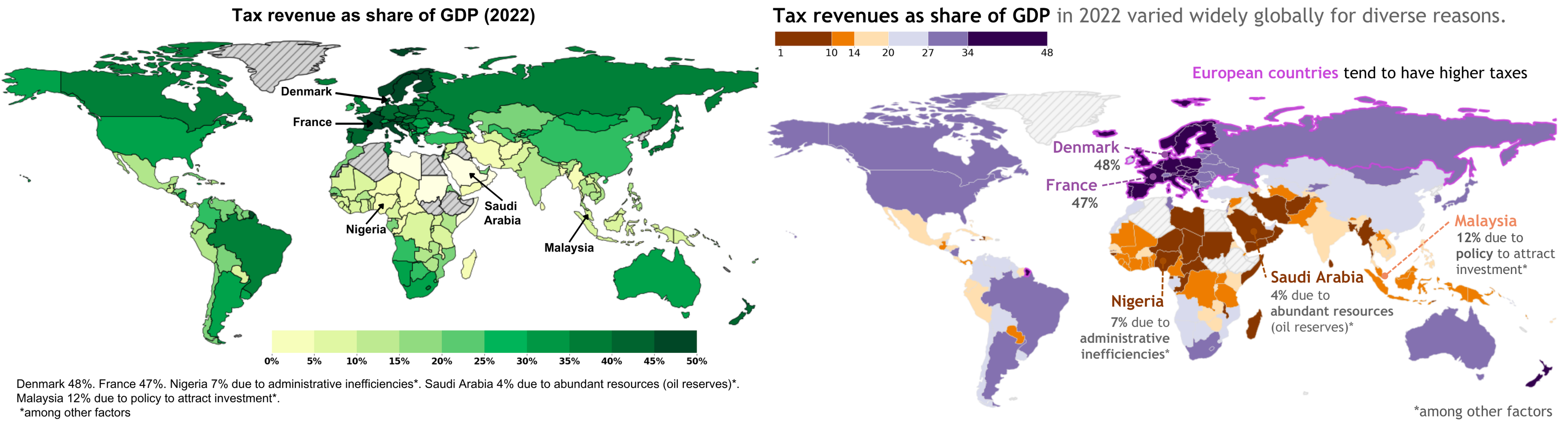}
    \caption{Visualisation \texttt{D} based on the topic of Tax Revenue as a share of GDP. The pair consists of the conventional visualisation \texttt{D-CV} [left] and the DS-enhanced visualisation \texttt{D-DS} [right].}
    \Description{A pair of maps each depicting the tax revenue as a share of GDP in 2022. The left visualisation is a traditional visualisation with a descriptive title, a legend, labels for particular country names, and a footnote paragraph highlighting key countries with reasons for their tax revenue percentage. The right visualisation is enhanced by data storytelling. It has an explanatory title and annotations labelling key countries along with with a reason for their tax revenue percentage.}
    \label{fig:visD}
\end{figure*}

The design process to create the pairs of visualisations for this study followed the \textit{Data Storytelling Framework} developed by \citet{zdanovic2022influence} and the visual enhancements described in the data storytelling and visualisation literature \cite{tufte2001visual, knaflic2015storytelling, ryan2016visual, echeverria2018exploratory, stokes2023striking}. \citet{zdanovic2022influence}'s framework is divided into three distinct phases – 1) \textit{Explore the data}, 2) \textit{Craft the visualisation}, and 3) \textit{Tell the story} – each with individual steps that are necessary to guide the visualisation creation process. While generating conventional visualisations generally involves going through the first two phases – data exploration and visualisation creation – generating data stories extends this process by performing a more nuanced crafting of the visualisation in the second phase and incorporating storytelling elements in the third phase, enriching the narrative and enhancing the communication of insights. The full details about the visualisation design, transformation process, and implementation, are available at \href{https://github.com/mmilesi45/DataStorytelling-ComplexTasks/blob/main/1.%20Visualisations/visualisation-design.pdf}{this link}. The resulting pairs of visualisations are depicted in Figures \ref{fig:visA}-\ref{fig:visD}. 

The conventional visualisations were crafted using the original aesthetics and visualisation types from the respective \textit{Our World in Data} data stories. To ensure a fair comparison between tasks solved with the conventional and DS-enhanced visualisations, we applied the same data filtering process (as outlined in \citet{zdanovic2022influence}) to include identical data points in both visualisations. Consequently, some conventional visualisations differ slightly from the original visualisation from \textit{Our World in Data}. For instance, in conventional visualisation A (Figure \ref{fig:visA}), certain European empires included in the original visualisation were excluded. Both the original \textit{Our World in Data} visualisations and the conventional visualisations from this study can be viewed via \href{https://github.com/mmilesi45/DataStorytelling-ComplexTasks/blob/main/1.%20Visualisations/visualisations.pdf}{this link}.

In the DS-enhanced visualisations, text from \citet{herre2023statecap} was adapted to the form of annotations and explanatory titles. Using the suggestions by \citet{kim2021towards} and \citet{stokes2023striking}, annotations were carefully placed to guide the audience towards the data points they were describing. Pre-attentive attributes for text, such as bolding particular terms or using colour to highlight to further draw attention towards key terms or data-points \cite{knaflic2015storytelling}. Additionally, a footnote was added to the bottom of the conventional visualisations to provide participants with the same contextual information, ensuring that differences in responses were due to how the information was presented (i.e., DS-enhanced vs. conventional) rather than the amount of information provided.

\subsection{Human-in-the-Loop-AI Question-Based Task Generation Process}
To generate question-based tasks aligned with the various levels of Bloom's taxonomy and ensure comprehensive coverage of the selected dataset, we adopted a \textit{Human-in-the-Loop-AI} collaborative process \cite{hao2024transforming}. This process involves instructing AI systems to generate question-based tasks on a large scale, followed by human review and refinement of the content \cite{hao2024transforming, attali2022interactive, zu2023automated}. Our question-based task generation process included four steps: 

\noindent \textbf{Step 1: Generation with LLMs.} Motivated by the latest research in automated item generation \cite{hao2024transforming, attali2022interactive, zu2023automated}, GPT-4 was used to create an initial pool of questions. The rationale behind using GPT-4 was to efficiently generate a large set of potential questions that aligned with both the content of the visualisations and the specific levels of Bloom's taxonomy. This aligns with emerging approaches, such as those used by Duolingo for generating assessment items, which leverage LLMs to efficiently create questions at varying levels of complexity while minimising human bias and keeping a consistent style and tone \cite{attali2022interactive}. For the first five levels of Bloom's taxonomy, the conventional visualisations were uploaded alongside specific prompts based on definitions from \citet{krathwohl2002revision}. The conventional visualisations were chosen to be uploaded as they contained the least amount of explanatory information to ensure fair comparisons. For example, to generate questions for the \textit{Understand} level, GPT-4 was instructed as follows: ``\textit{You are an expert learning scientist. Given the provided graph, generate 10 close-ended questions (e.g., multiple-choice, ranking questions, checkboxes) for the second level of Bloom's taxonomy: Understand. Determining the meaning of instructional messages, including oral, written, and graphic communication (e.g., Explaining)}''.

\noindent \textbf{Step 2: Validation with BloomBERT.}
A total of 89 questions were generated by GPT-4 which were then validated using BloomBERT \cite{BloomBERT}, a fine-tuned DistilBERT model on over 6K questions for 40 epochs achieving a classification accuracy of 91\% for Bloom's taxonomy levels. The 46 questions that BloomBERT accurately categorised into the intended Bloom's level were retained for further refinement. 

\noindent \textbf{Step 3: Adaptation by Authors.}
Following validation by BloomBERT, the resulting 46 questions were reviewed by two of the authors, who jointly refined them for clarity, relevance, and alignment with the visualisations. Low quality questions were discarded leaving 36 viable questions.

\noindent \textbf{Step 4: Validation by Human Annotators.}
Finally, the questions were validated by another two of the authors who were not involved in the previous steps. They categorised each question according to Bloom's taxonomy levels as described by \citet{burns2020evaluate} and \citet{krathwohl2002revision}. The inter-rater agreement between the human annotators was evaluated using Cohen’s Kappa, resulting in an initial agreement score of 0.89, indicating almost perfect agreement. Any questions where the annotators disagreed were discarded. An example of questions for visualisation \texttt{A} are available in appendix \ref{sec:appendix-questions}. 

\subsection{Study Design}
\begin{figure*}[htbp]
    \centering
    \includegraphics[width=1\linewidth]{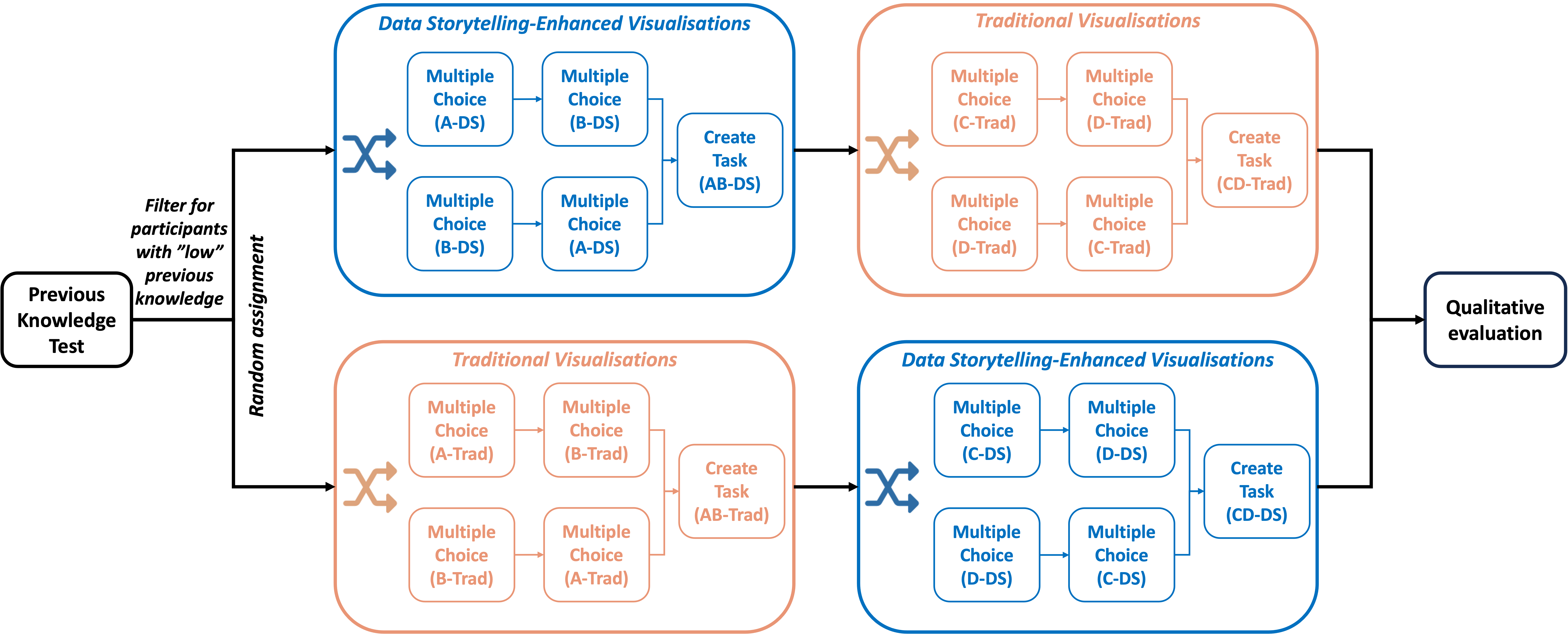}
    \caption{Outline of the study design comparing conventional and DS-enhanced visualisations. Participants, after a pre-test, were randomly assigned to view both types of visualisations and answer MCQ for each visualisation and an open task linking each pair. Lastly, participants provided a qualitative evaluation of their preferences.}
    \Description{This figure illustrates the structure of the controlled study. Participants first completed a knowledge pre-test to filter out those with prior knowledge on the topics. They were then randomly assigned to one of two groups: (1) viewing DS-enhanced visualisations for the first two visualisations and conventional visualisations for the last two (AB-DS, CD-CV) or (2) viewing the reverse order (AB-CV, CD-DS). Each participant completed multiple-choice tasks for both line charts (A and C) and maps (B and D), followed by a higher-order "Create" task where they integrated insights from both visualisations. Lastly, participants provided qualitative evaluations.}
    \label{fig:survey-structure}
\end{figure*}

We conducted a controlled study to investigate the impact of DS-enhancements in visualisations using a within-subjects design. Participants were exposed to both conditions: conventional visualisations and DS-enhanced visualisations. Figure \ref{fig:survey-structure} details the structure of the survey. The survey structure, managed via Qualtrics, handled the randomisation process to minimise order effects. The full questionnaire is available \href{https://github.com/mmilesi45/DataStorytelling-ComplexTasks/blob/main/Effects%20of%20Increasing%20Task%20Complexity%20on%20the%20Effectiveness%20of%20Data%20Storytelling%20Enhanced%20Visualisations%20-%20Questions.pdf}{here.}

\label{sec:instrument}

\noindent \textbf{Prior Knowledge Questions.} To ensure responses were based on information from visualisations rather than prior knowledge, inspired by \citet{zdanovic2022influence}, participants completed a self-assessment (five-point Likert scale) and a multiple-choice question on each topic. Participants who self-reported as \textit{Very Knowledgeable} or \textit{Extremely Knowledgeable} and answered correctly were excluded and compensated for their time with $1$ USD. 

\noindent \textbf{RQ1: Measuring Effectiveness \& RQ2: Efficiency.} Participants answered eight multiple-choice questions per visualisation, covering the first five levels of Bloom’s taxonomy, with an “\textit{I am not sure}” option to discourage guessing \cite{nobre2024,shao2024datastorytelling}. Visualisations \texttt{A} and \texttt{B} and \texttt{C} and \texttt{D} were presented alternately with and without DS enhancements (\texttt{AB-DS/CD-CV} or \texttt{AB-CV/CD-DS}), ensuring exposure to one conventional and one DS-enhanced line chart and map per participant. 

\noindent \textbf{RQ3: Measuring Quality of Response to Create Tasks.} Next, participants completed an open-ended "Create" task requiring them to integrate insights from visualisations. For example, participants wrote compositions combining information from \texttt{A} and \texttt{B} (e.g., \texttt{A-CV} and \texttt{B-CV}) and \texttt{C} and \texttt{D} (e.g., \texttt{C-DS} and \texttt{D-DS}).

\noindent \textbf{RQ4: Measuring Perception of Data Storytelling.} Lastly, participants compared DS-enhanced and conventional visualisations of the same type (e.g., line charts). They indicated preferences and justified their choices. Additionally, participants clicked on visual elements they found useful to simulate eye-tracking insights.

\subsection{Participants}
A priori power analysis using G*Power 3.1 determined that a sample size of 128 participants was needed to achieve sufficient statistical power ($\alpha = 0.05, \text{power} = 0.95$) using an effect size of 0.3 based on \citet{shao2024datastorytelling}. Three pilot studies preceded the main study: the first refined the survey design with feedback from four HCI experts; the second, involving five participants, estimated study duration; and the third, with 25 participants, tested randomisation and analysis. Participants who answered questions that changed during pilots (e.g., changing the heatmap questions to allow multiple clicks) had their results excluded from the final dataset.

Participants for the main study were recruited through CloudConnect. The study lasted a median of 51 minutes ($\text{IQR} = 30$), and participants were compensated $9$ USD. Pre-screening ensured fluency in English and a balanced sex ratio. Of 157 initial participants, 128 valid responses remained after excluding eight overly knowledgeable individuals, ten who failed attention checks, and 21 suspected of using generative AI for written tasks. Written responses with a calculated word-per-minute rate of over 50 \cite{dhakal2018observations} were manually checked for attributes that are characteristic of GPT-4, such as being overly verbose or formal in language \cite{kabir2024stack}.

The final sample included 63 males (49\%) and 65 females (51\%), aged 18–64 years ($\mu = 24.14, \sigma = 7.02$). Educational backgrounds varied: 41\% had some college education, 37\% held bachelor’s or associate degrees, and 14\% had high school diplomas. Participants were mostly from the US (109), with smaller numbers from the UK, Canada, Australia, New Zealand, and Ireland. Most participants (53–54\%) expressed confidence in understanding and interpreting data, with fewer than 10\% lacking confidence. Ethics approval was obtained from Monash University. Informed consent was secured from each participant.

\subsection{Metrics} 
\textit{Effectiveness} (RQ1) was measured as the percentage of correct responses, calculated separately for each visualisation type and cognitive category (e.g., \textit{Identify}). For a set \( S \), effectiveness was determined by averaging the binary correctness of responses, where each correct answer scored 1 and incorrect answers scored 0.

\textit{Efficiency} (RQ2) assessed the time participants took to answer questions correctly, measured from when a question was displayed to when the correct answer was submitted. Efficiency was calculated as the average time spent on correctly answered questions for each participant, separated by visualisation type and cognitive levels in Bloom’s taxonomy.

To compare text compositions across visualisation conditions, the \textit{Percentage Change} was used, reflecting the proportional difference in the number of sentences normalised by the larger value. This symmetric metric bounded changes between \(-1\) and \(1\). For example, the percentage change between DS-enhanced and conventional conditions was calculated as:

\[
\text{Percentage Change}_{\text{Overall}} = \frac{DS-enhanced - conventional}{\max(DS-enhanced, conventional)}\times 100
\]

\subsection{Data Analysis}
We used the Shapiro-Wilk W-test to evaluate the normality of the \textit{effectiveness} and \textit{efficiency} scores. The results indicated that these variables deviated significantly from a normal distribution, thus we used non-parametric tests for the analysis. Wilcoxon signed-rank test was used to compute differences between paired observations; and consequently used the matched pairs rank biserial correlation to calculate the effect size \cite{kerby2014simple}. We report the absolute value of the effect size. 

For the \textit{effectiveness} (RQ1) and \textit{efficiency} (RQ2), we calculated the score for each set of visualisations (DS-enhanced and conventional) for each participant, both overall and at each level of Bloom's taxonomy. We then applied the Wilcoxon signed-rank test, suitable for paired observations, to evaluate within-subject differences. This analysis was conducted both overall and across each level of Bloom's taxonomy, with corrections for multiple comparisons applied using the Benjamini-Hochberg (BH) procedure.

The \textit{Create} task (RQ3) involved sentence-level rhetorical structure analysis \cite{mann1987rhetorical}. Several frameworks could be employed for this analysis including Rhetorical Structure Theory \cite{burstein2003finding}, semantic content levels for accessible visualisations \cite{lundgard2022accessible} and Bloom’s taxonomy \cite{adar2021communicative,iqbal2023towards, iqbal2024towards}. Among these, we selected Bloom’s taxonomy for its hierarchical and systematic approach to evaluating cognitive processes involved in text comprehension and creation \cite{iqbal2023towards, iqbal2024towards} and its previous use as a taxonomy for visualisation learning objectives \cite{adar2021communicative,leerobbins2022learning, shao2024datastorytelling}.  The dataset consisted of 1818 sentences coming from 256 texts (two texts per participant). Four researchers developed a rubric for annotation (see Appendix \ref{sec:appendix-create-rubric}), with two independently labelling 32 sentences to validate the rubric, achieving an inter-rater reliability of 0.78 (Cohen's Kappa). Discrepancies were resolved through discussion, and remaining sentences were annotated individually, including Bloom’s taxonomy levels and uncertainty. The annotators then met to discuss the sentences where they were not certain to reach a consensus. 

The total number of sentences and their distribution across Bloom's levels were calculated separately for DS-enhanced and conventional conditions. Tasks from visualisation pairs (\texttt{AB-DS}/\texttt{CD-DS} and \texttt{AB-CV}/\texttt{CD-CV}) were aggregated to account for differences in visualisation sets, and potential effects of the presentation sequence, as explored in  \citet{hullman2013deeper}. A Wilcoxon signed-rank test, with Benjamini-Hochberg adjustments, was used to assess within-subject differences between conditions.

User preferences (RQ4) were evaluated based on the frequency of DS-enhanced visualisation selection (overall preference rate). Clickstream interactions generated heatmaps highlighting useful elements (e.g., titles, annotations).  In addition, we calculated the number of clicks on specific elements of focus including the title, legend, axis labels, specific data points within the graph, annotations on data points on the graph, and contextual information on data points or as a footnote. Thematic analysis of open-ended responses identified key themes, reviewed and verified by two researchers. Finally, the frequency of these themes was calculated for each group (DS-enhanced vs conventional).

    \section{Results}
In this section, we present results for each of our research questions.

\subsection{RQ1: Effectiveness}
\begin{figure}[t]
    \centering
    \includegraphics[width=1\linewidth]{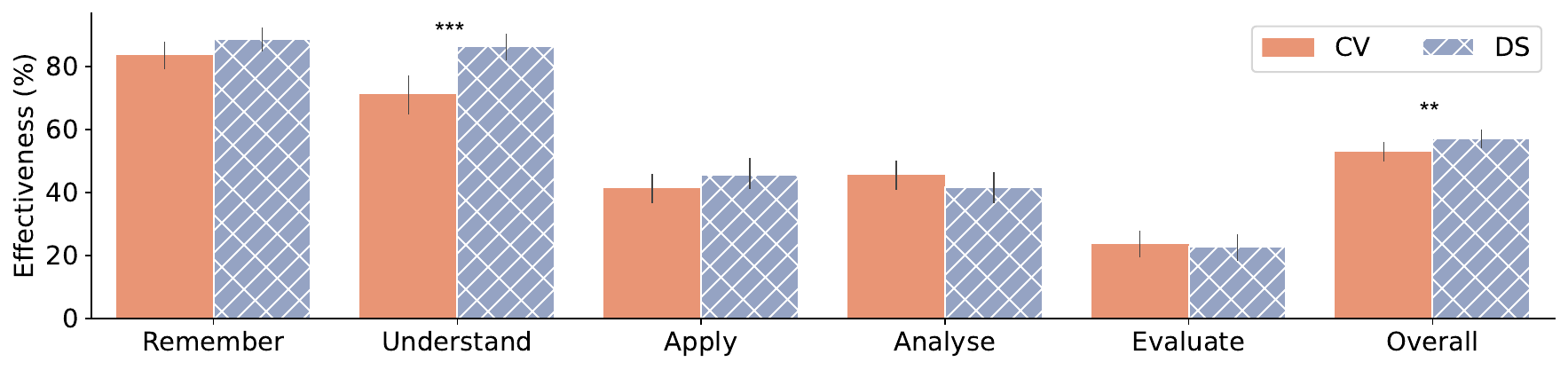}
    \caption{\textbf{\textit{Effectiveness}} of participants in answering questions across different cognitive levels of Bloom’s taxonomy for DS-enhanced and conventional visualisations. The effectiveness is measured as the percentage of correct answers. Error bars represent a 95\% confidence interval, $*$ indicates significance at $ p < 0.05 $, $**$ at $ p < 0.01 $, and $***$ at $ p < 0.001 $.}
    \label{fig:correctness}
    \Description{The figure displays a bar chart comparing the effectiveness (in percentage) of participants in answering questions across different cognitive levels of Bloom’s Taxonomy using either conventional (orange) or DS-enhanced (blue with crosshatch pattern) visualisations. The cognitive levels are listed on the x-axis as: Remember, Understand, Apply, Analyse, Evaluate, and Overall. The y-axis shows the effectiveness in percentage, ranging from 0 to 80.} 
\end{figure}

In the first analysis, we examined participants' \textit{effectiveness} when answering questions about the DS-enhanced and conventional visualisations for the first five levels of Bloom's taxonomy. As seen on the \textit{Overall} bar in Figure \ref{fig:correctness}, we observed that generally participants performed better in the DS-enhanced set of questions compared to the conventional set. The overall \textit{effectiveness} for the DS-enhanced and conventional visualisation sets were 57\% and 53\%, respectively. A Wilcoxon signed-rank test \footnote{The non-parametric test was performed after significant Shapiro tests, $W = 0.978$,  $p = 0.038$ and $W= 0.966$, $p = 0.003$} revealed a statistically significant difference with a moderate effect size ($W = 2217$, $p=0.004$,  $r =0.31$)

Transitioning to a more detailed analysis of the effectiveness in tasks of different complexity, Table \ref{tab:correctness_efficiency} and Figure \ref{fig:correctness} show the fine-grained effectiveness scores per Bloom's taxonomy level. In the first three levels: \textit{Identify}, \textit{Understand}, and \textit{Apply} -- the DS-enhanced condition outperformed the conventional visualisations. For \textit{Identify} ($W=667$, $p=0.96$, $r=0.22$) and \textit{Apply} ($W=1826$, $p=0.16$, $r=0.16$), the DS-enhanced effectiveness was 5\% higher than the conventional condition, though these differences were not statistically significant and the effect size was small. The most substantial difference occurred at the \textit{Understand} level ($W=336$, $p<0.001$, $r=0.59$), where the DS-enhanced effectiveness was 15\% higher than conventional visualisations, with a mean score of 86\% compared to 71\% for the conventional condition -- a statistically significant difference and a moderate effect. Different from the lower cognitive levels where DS-enhanced effectiveness was higher than the conventional scores, at the higher cognitive levels --\textit{Analyse} ($W=1398$, $p=0.17$, $r=0.13$) and \textit{Evaluate} ($W=954$, $p=0.70$, $r=0.05$) -- conventional visualisations performed slightly better by 4\% and 1\%, respectively, though these differences were not statistically significant.

\begin{table*}[t]
\centering
\caption{Comparison of \textit{Effectiveness} and \textit{Efficiency} between conventional and DS-enhanced visualisations at different cognitive levels. \textit{Effectiveness} is measured as percentage of correct answers and \textit{Efficiency} as the median correct answering time in seconds}
\begin{tabular}{l|ccc|ccc}
\toprule
\multirow{2}{*}{\textbf{Level}} & \multicolumn{3}{c|}{\textit{Effectiveness}} & \multicolumn{3}{c}{\textit{Efficiency}} \vspace{0.8mm}  \\ 
\cline{2-7}
 & conventional & DS-enhanced & \textit{p}-value & conventional & DS-enhanced  & \textit{p}-value \vspace{0.8mm} \\ \midrule
Identify & 84\% & \textbf{89\%} & 0.096 & \textbf{24.2} ($IQR$=21.9) & 27.4 ($IQR$= 19.9) & 0.661\\
Understand & 71\% & \textbf{86\%} & <0.001 & 18.6 ($IQR$=15.1) & \textbf{16.9} ($IQR$= 12.8) & <0.001 \\
Apply & 41\% & \textbf{46\%} & 0.156 & 32.7 ($IQR$=31.0) & \textbf{30.2} ($IQR$=22.6) & <0.001 \\
Analyse & \textbf{46\%} & 42\% & 0.145 & 39.3 ($IQR$=25.5) & \textbf{37.2} ($IQR$=27.6) & <0.001 \\ 
Evaluate & \textbf{24\%} & 23\% & 0.699 & 27.6 ($IQR$=26.5) & \textbf{27.4} ($IQR$=19.5) & <0.001 \\ 
Overall & 53\% & \textbf{57\%} & 0.004 & \textbf{29.7} ($IQR$=19.7) & 30.7 ($IQR$=20.5) & 0.249 \\ \bottomrule
\end{tabular}
\label{tab:correctness_efficiency}
\end{table*}

In summary, participants generally performed better with DS-enhanced visualisations, achieving a significantly higher overall effectiveness (57\%) compared to conventional visualisations (53\%). The most notable improvement for DS-enhanced visualisations was at the \textit{Understand} level, where effectiveness was 15\% higher than conventional visualisations

\subsection{RQ2: Efficiency}
\begin{figure}[h]
    \centering
    \includegraphics[width=1\linewidth]{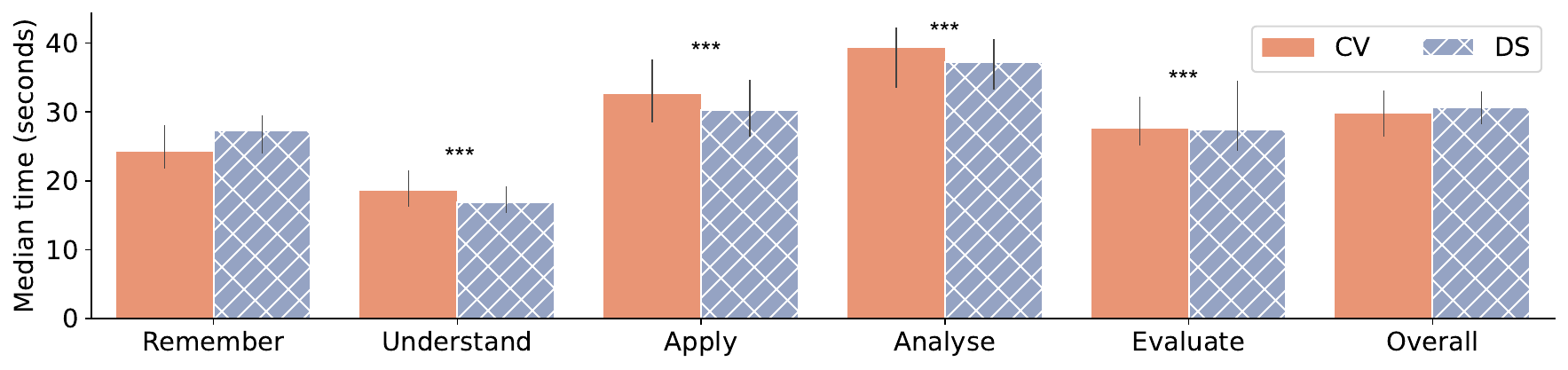}
    \caption{\textbf{\textit{Efficiency}} of participants in answering questions across different cognitive levels of Bloom’s Taxonomy for the DS-enhanced and conventional visualisations. The efficiency is measured as the average correct answer time. Error bars represent a 95\% confidence interval, $*$ indicates significance at $ p < 0.05 $, $**$ at $ p < 0.01 $, and $***$ at $ p < 0.001 $.}
    \Description{The figure displays a bar chart comparing the median time (in seconds) it took participants to answer questions across different cognitive levels of Bloom’s Taxonomy using either conventional (orange) or DS-enhanced (blue with crosshatch pattern) visualisations. The cognitive levels are listed on the x-axis as: Remember, Understand, Apply, Analyse, Evaluate, and Overall. The y-axis shows the median time in seconds, ranging from 0 to 40.} 
    \label{fig:efficiency-result}
\end{figure}

Regarding RQ2, we examined participants' \textit{efficiency} in answering the questions correctly. The efficiency data presented in Table \ref{tab:correctness_efficiency} and Figure \ref{fig:efficiency-result} indicates how quickly participants arrived at the correct answers in both sets of questions. The table shows the median seconds of participants and the IQR. In general, there was no significant difference in the overall \textit{efficiency} between the two visualisation methods ($W =3644$, $p=0.2$, $r=0.11$). However, interesting trends emerge at specific levels of Bloom's taxonomy. At the lowest level, \textit{Identify}, participants were able to identify the correct answers faster when using the conventional visualisations, with a median time of 24.2 seconds, which was more than 3 seconds faster than the median time for the DS-enhanced condition (27.4 seconds). Despite this, the within-subjects difference was not statistically significant ($W=3944$, $p=0.7$, $r=0.03$). 

In contrast, for the higher levels of Bloom's taxonomy --\textit{Understand} ($W=2471$, $p<0.001$, $r=0.18$), \textit{Apply} ($W=2482$, $p<0.001$, $r=0.02$), \textit{Analyse} ($W=2381$, $p=<0.001$, $r=0.04$), and \textit{Evaluate} ($W=691$, $p<0.001$, $r=0.16$) -- participants were more efficient when using the DS-enhanced visualisations. The median time per participant for the DS-enhanced visualisations was consistently lower than the conventional visualisations across these levels, and the differences within subjects were statistically significant. This suggests that participants were significantly more efficient at arriving at the correct answers when engaging with visualisations that included data storytelling elements at these higher cognitive levels.

To summarise, participants showed no significant difference in overall efficiency between DS-enhanced and conventional visualisations. However, participants were significantly more efficient with DS-enhanced visualisations at higher cognitive levels (\textit{Understand}, \textit{Apply}, \textit{Analyse}, and \textit{Evaluate}) with a low effect size.

\subsection{RQ3: Quality of Complex Writing Creation Tasks}
To investigate cognitive differences in the \textit{Create} task, we analysed participants' sentence-level responses. The focus was to determine whether visualisations with data storytelling elements (DS-enhanced) influenced participants' text composition compared to conventional visualisations (conventional). Overall, the average sentences for both conditions was 7.1 and there were no within-subject differences ($W = 1774$, $p = 0.958$, $r<0.01$). The most frequent cognitive level across conditions was \textit{Identify}, with participants writing an average of 2.6 sentences, followed by \textit{Analysis} (1.7 sentences) and \textit{Understand} (0.9 sentences). The least frequent levels were \textit{Evaluate} (0.7 sentences), \textit{Create} (0.4 sentences), \textit{Other} (0.36 sentences), and finally \textit{Apply}, which was the least frequent category with an average of 0.32 sentences.

\begin{figure}[h]
    \centering
    \includegraphics[width=1\linewidth]{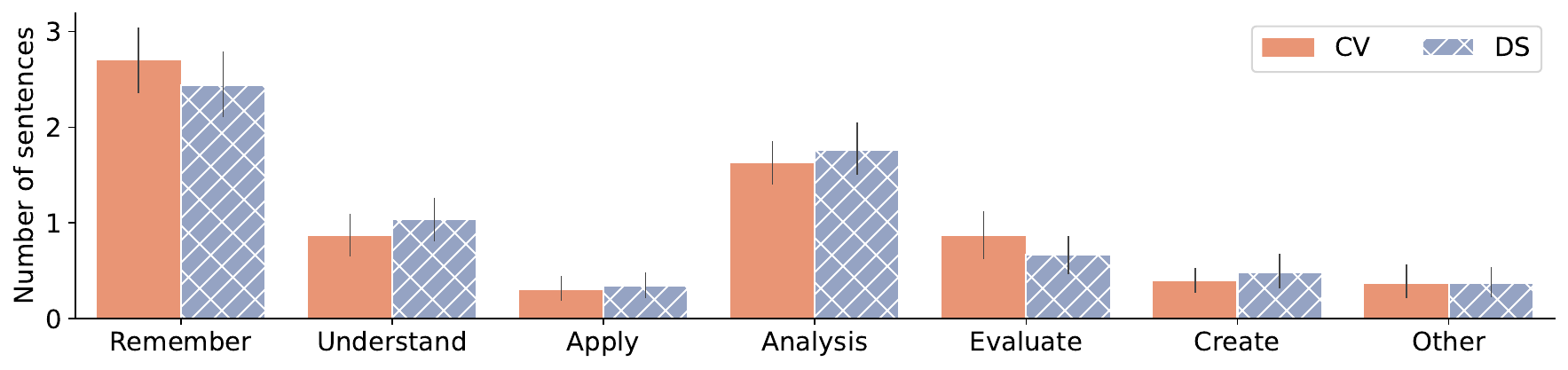}
    \caption{Distribution of sentences across different cognitive levels of Bloom’s taxonomy for DS-enhanced and conventional visualisations. Error bars represent a 95\% confidence interval.}
    \Description{The figure displays a bar chart comparing the number of sentences participants wrote on their Create task across different cognitive levels of Bloom’s Taxonomy using either conventional (orange) or DS-enhanced (blue with crosshatch pattern) visualisations. The cognitive levels are listed on the x-axis as: Remember, Understand, Apply, Analyse, Evaluate, and Overall. The y-axis shows the number of sentences, ranging from 0 to 3.} 
    \label{fig:create-dist}
\end{figure}
Figure \ref{fig:create-dist} visualises the distribution of sentences across cognitive categories for each condition. Analysis of the data, as shown in Table \ref{tab:create-levels}, revealed no significant within-subject differences in text composition between the DS-enhanced and conventional visualisations. However, some trends were notable: participants who viewed conventional visualisations tended to produce more sentences at the \textit{Identify} ($W=2152$, $p=0.19$, $r=0.15$) and \textit{Evaluate} level ($W=867$, $p=0.12$, $r=0.22$). 

\begin{table*}[t]
\centering
\caption{Comparison of the average number of sentences participants wrote across cognitive levels when using DS-enhanced and conventional visualisations.}
\begin{tabular}{l|ccccc}
\toprule
\textbf{Level} & DS-enhanced & conventional & \texttt{DS - CV} & Percentage Change (\%) & \textit{p}-value \vspace{0.8mm} \\ \midrule
Identify   & 2.44 & \textbf{2.70} & -0.26 & -11\% & 0.19 \\
Understand & \textbf{1.04} & 0.86 & \textbf{0.18}& 21\% & 0.10 \\
Apply      & \textbf{0.34} & 0.30 & \textbf{0.04}& 16\% & 0.57 \\
Analysis   & \textbf{1.77} & 1.63 & \textbf{0.14}& 9\% & 0.55 \\
Evaluate   & 0.66 & \textbf{0.86} & -0.20 & -29\% & 0.12 \\
Create     & \textbf{0.48} & 0.39 & \textbf{0.09} & 24\% & 0.39 \\
Other      & 0.37 & 0.36 & 0.01 & 2\% &0.99 \\
\bottomrule
\end{tabular}
\label{tab:create-levels}
\end{table*}

As seen in Table \ref{tab:create-levels}, \textit{Evaluate} had the largest percentage change between the conditions involving DS-enhanced and conventional visualisations. When the participants composed a text using the conventional visualisations, they wrote on average 29\% more \textit{Evaluate} sentences. In contrast, those who used DS-enhanced visualisations wrote 21\% more sentences in the \textit{Understand} level ($W=1188$, $p=0.1$, $r=0.21$), 9\% more at the \textit{Analysis} level ($W=1818$, $p=0.55$, $r=0.07$), and 24\% more at the \textit{Create} level  ($W=622$, $p=0.39$, $r=0.13$). 

To sum up, although not statistically significant, the most notable differences between the tasks with conventional and DS-enhanced visualisations were that participants using DS-enhanced wrote, on average, 21\% more \textit{Understand} sentences and 24\% more \textit{Create} sentences, but 29\% fewer \textit{Evaluate} sentences

\subsection{RQ4: User Perception}

To explore how visualisations with data storytelling elements are perceived compared to conventional visualisations, we analysed participants' preferences and justifications. The results showed that the DS-enhanced visualisations were favoured over the conventional visualisations 60\%  of the time across both graph types. 

When participants chose the DS-enhanced visualisation over the conventional visualisation (60\% of the time), we found that in the justification of their choice, the most commonly mentioned element was the placement of contextual information directly next to relevant data points (e.g., next to country names) cited 84\% of participants. In particular, one participant stated: \textit{``I think the blurbs of text that explain the trends in the lines give some much-needed context in a visually pleasing format''}. Another noted that the DS-enhancements permitted them to ``\textit{piece data connections together like a puzzle}''. Additionally, 17\% specifically mentioned annotations on data points on the graph, specifically the display of exact values (such as percentages) on the visualisation. One participant noted the direct percentages: \textit{``The percentages were up with the country name and made it easier to process, requiring less time to find percentages''}. Furthermore, 16\% of participants mentioned that the high contrast between colours, particularly among different categories, made distinctions clearer; while 11\% mentioned the different dashed line patterns in the line charts, for example, one participant stated: \textit{``I think the difference in the pattern of the lines in the second graph also made the countries more differentiable''}. Lastly, 5\% praised the focus on important areas. As one participant stated: \textit{I appreciate that it focused on specific places and didn't color the other ones that were not of focus — it helped keep the perspective and focus on the areas of interest instead of being overwhelming''}. 

In contrast, when participants preferred the conventional visualisation (40\% of the time), 56\% of the justification cited a preference for minimal clutter, favouring charts that were not overloaded with information. Furthermore, 21\% referred to the map's legend preferring a legend where the intervals between different categories are consistent and evenly distributed and lastly, 13\% expressed a preference for a monotonic colour scale.

\begin{figure}[t]
    \centering
    \begin{minipage}{0.49\textwidth}
        \centering
        \includegraphics[width=\linewidth]{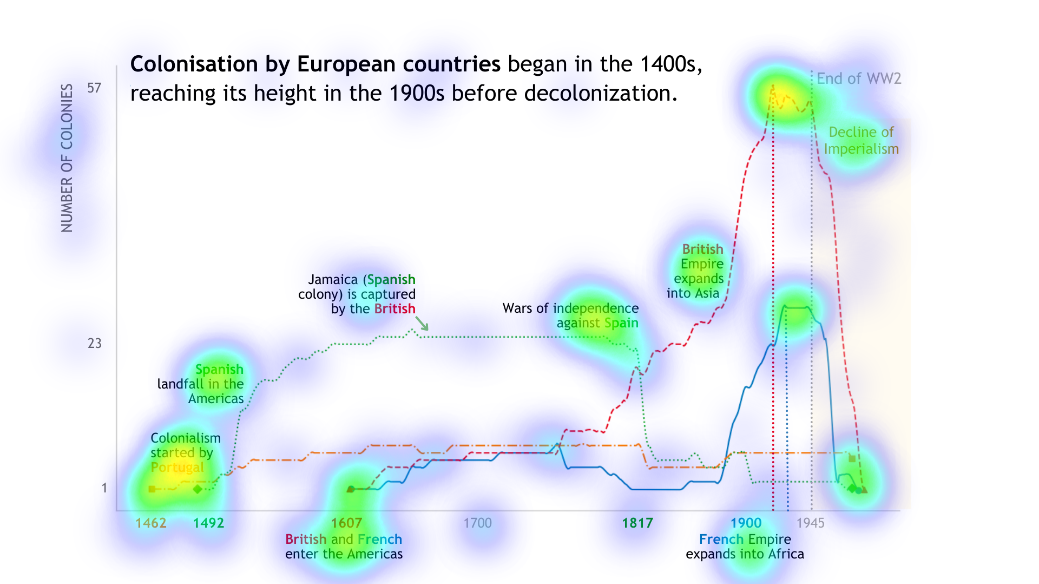}
        \caption{Heatmap showing the aspects that participants found useful in \texttt{A-DS}.}
        \label{fig:heatmap-A}
        \Description{Heatmap displaying the useful elements clicked in the DS-enhanced line chart (A-DS). Participants clicked most on annotations describing the trends over time, as well as key historical events marked on the graph.}
    \end{minipage}
    \hfill
    \begin{minipage}{0.49\textwidth}
        \centering
        \includegraphics[width=\linewidth]{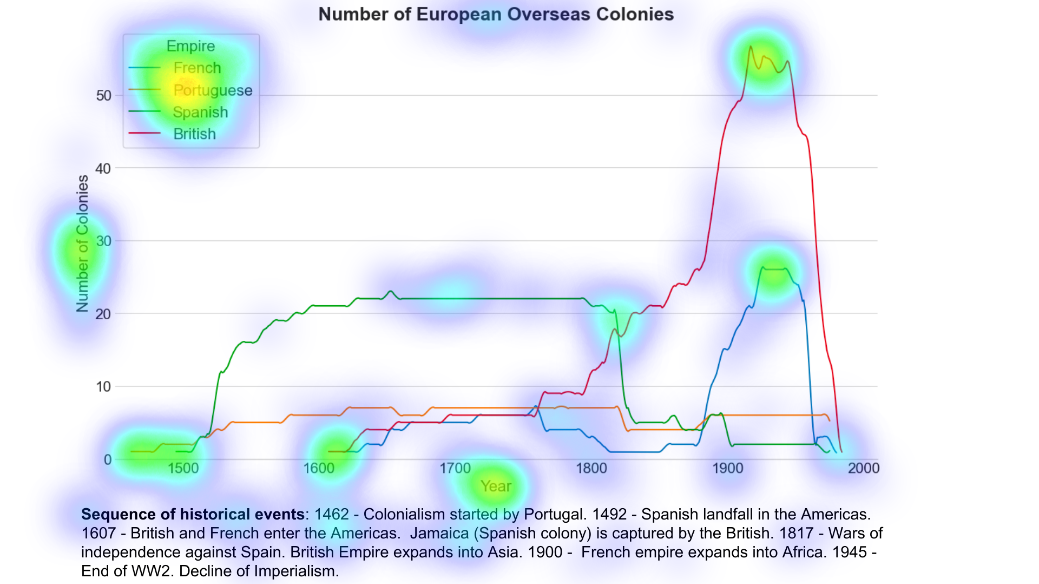}
        \caption{Heatmap showing the aspects that participants found useful in \texttt{A-CV}.}
        \Description{Heatmap depicting the useful elements clicked in the conventional line chart (A-CV). Participants highlighted the legend and y-axis labels, showing less interaction with data points compared to the DS-enhanced version.}
        \label{fig:heatmap-A-trad}
    \end{minipage}
\end{figure}

\begin{figure}[t]
    \centering
    \begin{minipage}{0.49\textwidth}
        \centering
        \includegraphics[width=\linewidth]{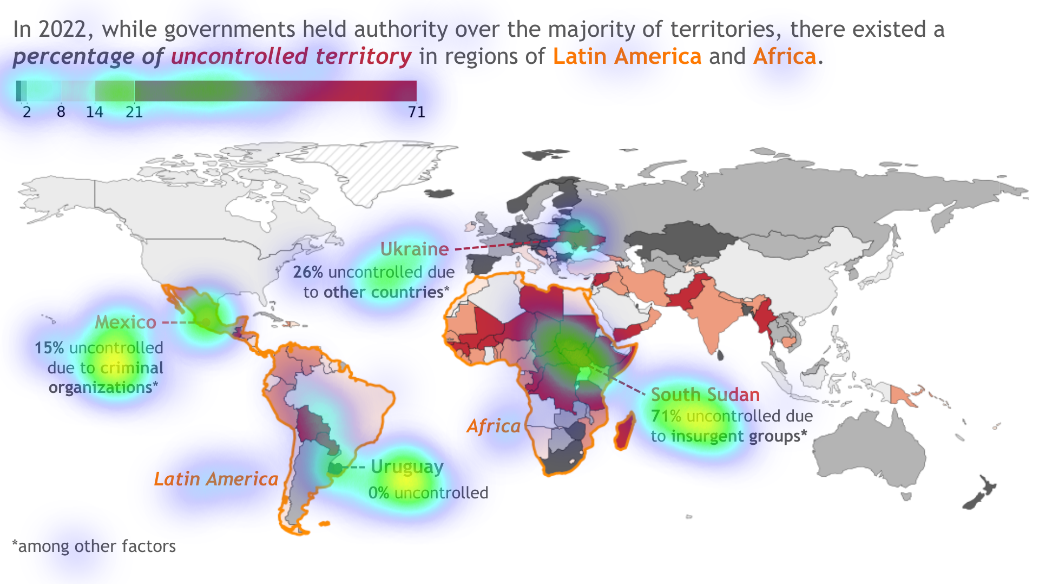}
        \caption{Heatmap showing the elements that participants found useful in \texttt{B-DS}.}
        \Description{Heatmap showing the useful elements clicked by participants in the DS-enhanced map (B-DS). The most clicked areas include annotations on data points, such as country names, and key information presented within the map.}
        \label{fig:heatmap-map-C}
    \end{minipage}
    \hfill
    \begin{minipage}{0.49\textwidth}
        \centering
        \includegraphics[width=\linewidth]{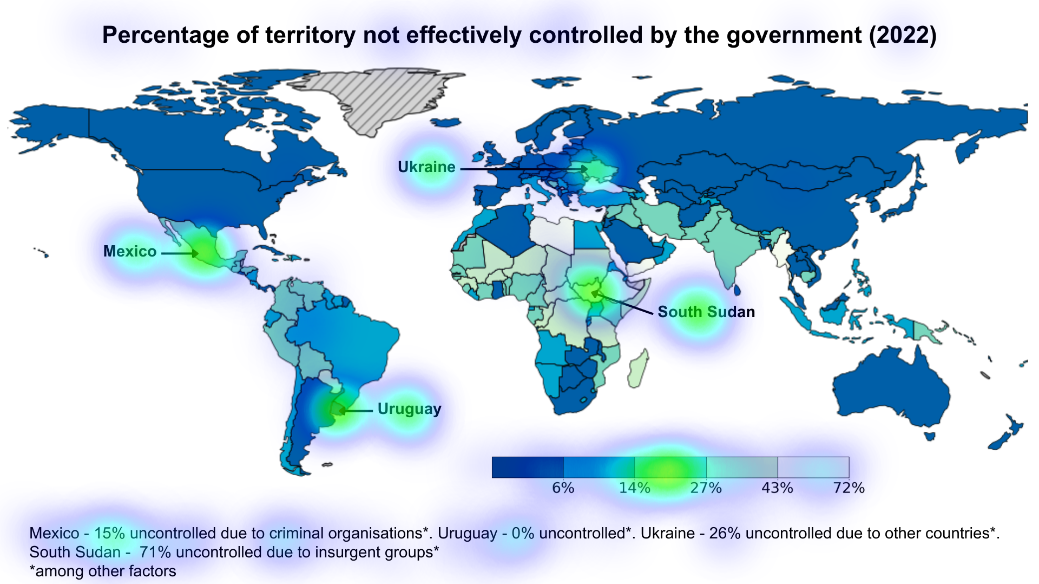}
        \caption{Heatmap showing the aspects that participants found useful in \texttt{B-CV}.}
        \label{fig:heatmap-map-D}
        \Description{Heatmap illustrating the useful elements clicked by participants in the conventional version of the map (B-CV). The key elements clicked include the legend and some country names, with fewer clicks on specific data points compared to the DS-enhanced version.}
    \end{minipage}
\end{figure}

Complementing participants' written justifications, Figure \ref{fig:heatmap-map-C} and Figure \ref{fig:heatmap-map-D} display heatmaps of participants' clicks for visualisation \texttt{B}. In both versions, conventional (Figure \ref{fig:heatmap-map-D}) and DS-enhanced (Figure \ref{fig:heatmap-map-C}), the most popular areas were the annotations on data points on the graph (like the country names) and the legends. Moreover, the heatmaps shown in Figure \ref{fig:heatmap-A} (DS-enhanced) and Figure \ref{fig:heatmap-A-trad} (conventional) show the areas perceived as useful from visualisation \texttt{A}. 

In Figure \ref{fig:heatmap-A} (DS-enhanced) the contextual information on data points were clicked on 170 times whereas in Figure \ref{fig:heatmap-A-trad} (conventional) the contextual information as a footnote received three times fewer clicks (only 46 clicks). Instead, in Figure \ref{fig:heatmap-A-trad} (conventional), the most highlighted element is the legend which was clicked on 75 times. Interestingly, both charts have the same y-axis label yet it is one of the main chosen elements in Figure \ref{fig:heatmap-A-trad} (conventional) clicked on it by 29 participants and it was less frequently chosen in Figure \ref{fig:heatmap-A} (DS-enhanced) where it was only selected by 8 participants.

In summary, participants generally favoured DS-enhanced visualisations over conventional ones 60\% of the time, citing the placement of contextual information next to relevant data points (84\%), clear colour contrast (16\%), and distinct line patterns (11\%) as key reasons. In contrast, those who preferred conventional visualisations (40\%) valued minimal clutter (56\%), evenly spaced legends (21\%), and monotonic colour scales (13\%).
    \section{Discussion}
In this section, we present a summary of the key findings per research questions, provide the implications for both research and practice, and address the limitations of our study. Additionally, we outline potential directions for future research.

\subsection{Summary of Results and Research Questions}
\subsubsection{RQ1: Effectiveness} Participants demonstrated higher \textit{effectiveness} scores on the DS-enhanced visualisations compared to the conventional visualisations, with a statistically significant difference and a small effect size. Interestingly, the \textit{effectiveness} scores were higher for the first three levels of Bloom's taxonomy—\textit{Identify}, \textit{Understand}, and \textit{Apply}—when using DS-enhanced visualisations. These results are consistent with those of \citet{shao2024datastorytelling}, who also observed that DS significantly improved participants' ability to identify and comprehend key data insights in the first two levels of cognition. However, unlike previous work \cite{shao2024datastorytelling, zdanovic2022influence}, we studied the effectiveness at further levels of Bloom's taxonomy. For tasks requiring higher-order cognitive skills, such as \textit{Analyse} and \textit{Evaluate}, the \textit{effectiveness} scores were higher for the conventional visualisations. Although the differences were not significant, this may suggest that DS elements are most effective in simpler tasks, where narrative elements help structure information, but less so when participants must engage in more abstract thinking, where the narrative may not provide additional value or could even distract from critical thinking processes. The more structured nature of DS-enhanced visualisations, which often highlight specific data points, might direct users' attention and thinking, whereas conventional visualisations offer more flexibility for individual interpretation. 

\subsubsection{RQ2: Efficiency} Regarding \textit{efficiency}, participants showed no significant overall difference between DS-enhanced and conventional visualisations. However, at higher cognitive levels (\textit{Understand}, \textit{Apply}, \textit{Analyse}, \textit{Evaluate}), participants were significantly more efficient with DS-enhanced visualisations while for conventional visualisations they were slightly quicker at the \textit{Identify} level. Our findings from the \textit{Identify} level are in line with \citet{shao2024datastorytelling} where at tasks of lower cognitive levels, like \textit{Identify}, the DS elements on visualisations made participants take longer to answer questions correctly. One possible explanation is that the additional narrative and contextual elements introduced in DS-enhanced visualisations can distract or delay participants when the task requires simple recall or recognition. In contrast, conventional visualisations, being more straightforward and minimalistic, allow for quicker retrieval of known information. 

Another possible explanation may be related to the hierarchical ordering of the tasks. All the participants were presented tasks following the order Bloom's taxonomy, meaning that the first question was an \textit{Identify} question followed by \textit{Understand} and higher cognitive levels. When participants were exposed to the DS-enhanced visualisations, they may have taken longer in the \textit{Identify} tasks than in the conventional visualisation condition as they were reviewing all of the information available in the visualisation. Once they had understood the story they were faster at answering correctly the questions from all the higher levels (\textit{Understand}, \textit{Apply}, \textit{Analyse}, \textit{Evaluate}). This is consistent with the discussion from \citet{liu2020survey} and \citet{galati2021} who suggest that individuals will take time to engage in sensemaking behaviours with visualisations in order to understand the contents before they complete any tasks using the visualisation's information. The findings from this study also resonates with the argument put forth by \citet{ryan2016visual} who believes that individuals will typically notice visual attributes, such as colour, before focusing on the deeper meaning of the visualisation. It is possible that participants took the time to perceive the artistic visual elements of the DS-enhancements before reviewing the intended message of the data story.

\subsubsection{RQ3: Quality of Complex Creation Tasks} Participants' text composition showed no significant overall difference between DS-enhanced and conventional visualisations. However, conventional visualisations prompted more sentences at the \textit{Identify} and \textit{Evaluate} levels, while DS-enhanced visualisations led to more sentences in the \textit{Understand}, \textit{Analyse}, and \textit{Create} categories. The pattern of engaging in different levels of cognition depending on the added elements on the visualisation resembles the findings from \citet{stokes2023striking}, where participants engaged in deeper semantic content takeaways when the text was placed near relevant data points on the chart, rather than on the title.  In our study, contextual information on data points in the DS-enhanced visualisations may have contributed to participants engaging more with higher-level cognitive tasks, providing further evidence of the relationship between text placement and cognitive depth in visualisation-based tasks.

One possible explanation for the higher number of \textit{Identify} sentences in the conventional visualisation condition is that participants may have felt the need to restate and emphasise information that was less visually highlighted, as opposed to DS-enhanced where key values were already emphasised. This could have led participants in the DS-enhanced condition to move beyond simple recall and engage more frequently at the \textit{Understand} level and reach higher cognitive levels, such as \textit{Apply}, \textit{Analyse}, and \textit{Create}. Interestingly, participants using conventional visualisations produced more \textit{Evaluate} sentences. Similar to the discussion on RQ1, this could be attributed to conventional visualisations being more open-ended, thus encouraging interpretation and critical thinking. In contrast, DS-enhanced visualisations, which are designed to guide users toward a specific message or narrative, may reduce the need for critical evaluation, such as identifying missing elements that could influence interpretation. This hypothesis aligns with the work by \citet{kong2018} on the effect of visualisation titles which found that participants were often unaware of the potential bias in visualisations leading to misconceptions of information. Our work extends their findings suggesting that data storytelling elements make the visualisation more convincing and there is less awareness of the limitations or biases of the information presented and the way it is presented. \citet{rogha2024impact} hypothesised that the passive data interpretation could be due to the cognitive load induced by the narrative elements.  In the work of \citet{garreton2024attitudinal}, visualisations are effective in eliciting emotional impact and driving attitude change. Building on this, we argue that while the persuasive power of such visualisations can be instrumental in driving action, such as addressing urgent issues like the water crisis \cite{garreton2024attitudinal}, this same persuasiveness may have unintended consequences. Specifically, these visualisations could inadvertently limit users’ ability to question the presented narrative, potentially leading to over-reliance on the guided interpretation provided.

\subsubsection{RQ4: User Perception} Lastly, regarding user preferences, the majority of participants preferred DS-enhanced over conventional visualisations for the placement of contextual information and clear distinctions between countries, while conventional visualisations were preferred for their simplicity and minimal clutter. This aligns with the findings by \citet{yen2020} where participants perceived visualisations as useful when the text was presented in shorter segments that focused on one aspect at a time rather than all together. The idea that DS enhancements, such as annotations and explanatory titles, help to contextualise information was shared by the participants from \citet{echeverria2018exploratory, stokes2022text, milesi2024storytelling} who praised textual elements for their ability to make insights understandable and accessible. Similarly, \citet{stokes2023striking} found that participants preferred charts with a narrative or story to a text-only version of the data or visualisations without text annotations. A limitation of their work was the use of synthetically generated line charts, which lacked the complexity of real-world examples. Our study addresses this gap by providing empirical validation of \citet{stokes2022text,stokes2023striking} observations with real-world datasets, particularly emphasising the importance of placing contextual information near relevant data points. Additionally, the results are congruent with \citet{bateman2010} where visualisations with DS elements were perceived as more attractive, enjoyable, and easier to read and remember. Lastly, it is worth noting that there are individual preferences for information presentation or omission \cite{lundgard2022accessible}. \citet {lundgard2022accessible} advocate for adaptable descriptions that balance elemental, perceptual, and contextual information to meet diverse user needs in natural language descriptions of visualisations. We propose that this principle also applies to contextual annotations in DS-enhanced visualisations.

\subsection{Implications for Research}
In the literature, DS-enhancements have been a focal point for their perceived ability to facilitate the sensemaking process and more effectively and efficiently guide individuals towards complex insights compared to conventional data visualisations \cite{segel2010narrative, boy2015storytelling, Daradkeh2021}. This study, along with the work of \citet{shao2024datastorytelling}, has suggested that with respect to effectiveness, this idea holds for tasks that require a low level of cognitive processing. However, our work indicates that for tasks that require higher order thinking DS-enhancement may be less effective at facilitating the sensemaking process such that individuals can uncover patterns and insights on their own \cite{campos2021makingsense}. This indicates that the complexity of a task can significantly impact the \textit{effectiveness} of DS-enhanced visualisations. This presents an interesting direction for future works in establishing whether the nature of the task (e.g., written, verbal, etc.) can impact the effectiveness just as much as the complexity of the task.

In terms of efficiency, while \citet{shao2024datastorytelling} observed no improvements in speed between DS-enhanced and conventional visualisations, this study indicated that DS-enhancements can significantly impact how quickly individuals are able to extract insights from complex data. This is consistent with the findings from \citet{echeverria2018exploratory, de2014evaluating}, and \citet{Pozdniakov2023how} who argue that DS-enhancements enable a deeper understanding of the presented narrative and thus facilitate faster interpretation of the material. This research could be expanded to investigate the specific DS-enhancements that aid in efficient extraction of insights or examine whether the improvements in efficiency are also observed different storytelling genres \cite{segel2010narrative}, such as data videos \cite{amini2015datavideos, xian2023datavideos} or data comics \cite{bach2017emerging, milesi2024comics}.

This paper contributes to the InfoVis and HCI literature by building on prior frameworks, such as those proposed by \citet{adar2021communicative} and \citet{shao2024datastorytelling}, to systematically assess the role of DS in supporting cognitive tasks across the full spectrum of complexity, as classified Bloom's taxonomy. Building on this theoretical foundation, we demonstrate its applicability to real-world tasks requiring both lower-order and higher-order skills through a large-scale empirical evaluation of DS-enhanced visualisations. Our study spans all levels of Bloom’s taxonomy, and provides insights into how DS can support higher-order cognitive levels -- \textit{Apply} to \textit{Create} -- which have been underexplored in the literature but are critical for communicating complex ideas in practice \cite{morth2023scrollyvis, seyser2018scrollytelling}. Unlike the work of \citet{leerobbins2022learning}, which focuses primarily on design specifications and user preferences, this study evaluates how these design principles translate into measurable improvements in task performance across varying levels of cognitive complexity. Additionally, we introduce a human-in-the-loop-AI task generation process, which combines large language models, domain-specific validation tools, and human oversight to create tasks aligned with Bloom's taxonomy. This approach provides a framework for designing cognitively diverse tasks, and similar to \citet{yan2024data,yan2024vizchat}, highlights the potential for human-AI collaboration in the HCI and InfoVis community.

\subsection{Implications for Design and Practice}
The different effects of DS-enhanced and conventional visualisations across various cognitive levels present distinct challenges and opportunities for researchers, practitioners, and designers. DS elements can significantly improve the effectiveness \textbf{(RQ1)} in understanding tasks and improve the efficiency \textbf{(RQ2)} in higher-order cognitive tasks requiring interpretation and critical thinking. Educators and trainers could consider adding DS elements based on specific learning objectives and the expected cognitive engagement levels of their audience. This could lead to higher learning efficiency and effectiveness, particularly in data-driven disciplines.

Despite the advantages of DS-enhanced visualisations in improving effectiveness in the \textit{Understanding} tasks and efficiency across most cognitive levels, it is noteworthy that in the \textit{Create} task, where users composed text based on the visualisations, participants using DS-enhanced produced fewer \textit{Evaluate} sentences (\textbf{RQ3}). Although this difference was not statistically significant, it warrants further investigation. This finding suggests a potential risk of DS-enhanced visualisations in contexts such as news or sensitive topics, where presenting a single narrative may inadvertently discourage critical thinking and reduce users' inclination to question underlying assumptions or consider missing information not presented in the graph. Understanding this effect more deeply could highlight an important limitation of data storytelling, especially when fostering critical evaluation is essential \cite{srinivasan2019augmenting}. This is consistent with the concerns raised by \citet{milesi2024storytelling}, who discussed the potential for critical data points or narratives to be inadvertently obscured by the inclusion of DS enhancements.

Based on the results of this study, we recommend the following:
\begin{itemize}
    \item For complex cognitive tasks, DS-enhancements did not support the correctness (\textbf{RQ1}) but tended to facilitate more \textit{efficient} information retrieval (\textbf{RQ2}). Consequently, conventional visualisations should be enhanced with DS elements in scenarios where the primary objective is to complete complex tasks more efficiently. 
    \item The relatively low level of critical evaluation in the DS-enhanced condition compared to the conventional visualisations (\textbf{RQ3}) suggests that while DS-enhancements can be leveraged to guide attention, they may lead to individuals interpreting the data more passively compared to those who explore the visualisations and discover the insights themselves. Therefore, depending on the context, it is important to balance the implementation of DS-enhancements to ensure that the intended message is still clear while also permitting critical evaluation of the data. For example, by using explicit elicitation techniques like contrastive narratives \cite{rogha2024impact} or interactive elements \cite{morth2023scrollyvis,shi2023breaking}. 
    \item Given that not all participants preferred DS-enhanced visualisations over conventional visualisations (\textbf{RQ4}), designers should consider creating adaptable visualisation tools that can modify the level of narrative complexity and data storytelling elements based on user preferences \cite{liu2020survey}, feedback, or the intended cognitive load of the task \cite{carenini2014highlighting}. As noted by \citet{figueiras2014narrative} if a user is proficient in the task, DS-enhanced visualisations may be perceived as ``\textit{boring}''. Conversely, users who are not proficient in the task may find highly exploratory visualisations difficult to engage with or comprehend \cite{figueiras2014narrative}. Therefore, the integration of user control features, such as the ability to toggle between more and less complex narrative features, may help accommodate a wider range of users and use cases.
\end{itemize}

\subsection{Limitations and Future Works}
A limitation of this study is the focus on line charts and maps as visualisation types. This choice was made to minimise potential fatigue due to the survey's length \cite{fereshteh2018evaluating} and to maintain consistency with the real-world context of \textit{Our World in Data}, whose preexisting data-driven narratives constrained the design to align with the information and presentation of the original visualisations. As a result, factors such as variations in contextual annotations (explored in \citet{stokes2023striking}) or DS-enhancements were not comprehensively explored. Instead, this study prioritised observing differences in task-solving performance across varying complexities within a defined DS-enhancement framework. Future studies should expand on these findings by exploring a broader range of visualisations (e.g., bar charts, scatter plots) and storytelling components. This study validates findings from studies that used synthetic data (e.g., \citet{stokes2023striking}) with real-world visualisations, serving as a foundation for future research to explore the generalisability of these results in more realistic contexts, such as full articles \cite{rogha2024impact} or interactive narratives \cite{shi2023breaking,morth2023scrollyvis}. For example, including breaking the fourth wall techniques, such as the \textit{Golden Hook} for capturing interest and the \textit{Magic Mirror} for tailoring insights as done in \citet{shi2023breaking}; as well as visual narrative flow elements including story layout and progression \cite{mckenna2017visual}, or sequencing techniques \cite{hullman2013deeper}.

While participants were instructed to not use external resources to answer questions and ``\textit{I am not sure}'' was given as an option for the multiple-choice questions, the online nature of this study means that it is possible that participants used search engines or GenAI to answer the questions. Despite efforts to exclude participants suspected of using GenAI in the written tasks, based on indicators such as words per minute, specific vocabulary, and grammatical structures, it remains difficult to completely verify the use of external sources. Consequently, conducting this study in a controlled in-person setting could mitigate the risk of cheating and obtain a richer qualitative dataset as participants work through the questions.

Another limitation of the study is the open-ended nature of the \textit{Create} task. No specific instructions were provided to participants outside of the general prompt asking them to write about the relationship between the visualisations. This left both the interpretation of the question and the content of their response to the discretion of the participants. As there is variation in the participants' demographics (age, background, country of origin, etc.) this may have influenced the responses to RQ3.

    \section{Conclusion}
This study compared data storytelling enhanced visualisations to conventional visualisations across tasks of varying complexity based on Bloom’s taxonomy. In particular, we focused on line charts and choropleth maps to communicate the data stories and support task completion. The results show that the effectiveness and efficiency are not significantly affected in the simpler tasks like extracting values from the graphs (\textit{Identify}). Whereas, it does improve the effectiveness in \textit{Understanding} tasks and the efficiency in higher-thinking tasks ranging from \textit{Understanding} to \textit{Evaluating}. Interestingly, DS-enhanced visualisations reduced the number of \textit{Evaluate} sentences in creation tasks, suggesting that they may decrease critical thinking by making users overly confident in the presented data. This highlights potential risks when using DS-enhanced visualisations in sensitive contexts, such as news or policy discussions, where critical evaluation is crucial. 
Overall, most participants preferred DS-enhanced visualisations for their clarity and contextual information, while some favoured the simplicity of conventional visualisations. These findings suggest that visualisation design should be adaptable to user preferences, task complexity, and task goals.

\bibliographystyle{ACM-Reference-Format}
\bibliography{reference.bib}

    \newpage
\appendix
\onecolumn

\section{Example Questions Aligned with Bloom's Taxonomy}
\label{sec:appendix-questions}

\begin{table}[h!]
\centering
\small
\caption{Examples of questions aligned with Bloom's Taxonomy levels used in the study. The examples are drawn from visualisation \texttt{A}, with questions ranging from multiple-choice formats to open-ended tasks for the \textit{Create} level.}
\begin{tabular}{p{3.4cm}p{4.4cm}p{6.5cm}}
\toprule
\textbf{Level \& Definition} & \textbf{Example Question} & \textbf{Possible Answers} \\ 
\midrule
\textit{Identify}: Retrieving relevant knowledge (e.g., Recognising) &Which Empire had overseas colonies as early as the 1460s, as shown in the graph? & French; Spanish; \textbf{Portuguese}; British \\ 
\midrule
\textit{Understand}: Determining the meaning of instructional messages, including oral, written, and graphic communication (e.g., Explaining) & Which is the most probable explanation for the observed trend in the Spanish Empire line from 1800-1850? & Increased military conquests by Spain in Africa; \textbf{Wars of independence in Latin America}; Napoleonic Wars that led to the cession of colonies to France; The liberal principles in the 1812 Constitution were implemented throughout the Spanish Empire \\ 
\midrule
\textit{Apply}: Carrying out or using a procedure in a given situation. (e.g.,Implementing) & How would the graph change if the British Empire had captured half of Spain's colonies in 1700? Select all that apply. & \textbf{The distance between the French plot line and the British plot line would increase in 1700}; The distance between the British plot line and the Spanish plot line would decrease in 1700; The British plot line would reach its maximum in 1700; The Portuguese plot line would be greater than the Spanish plot line in 1700; The distance between the Portuguese plot line and the Spanish plot line would increase in 1700 \\ 
\midrule
\textit{Analyse}: Breaking material into its constituent parts and detecting how the parts relate to one another and to an overall structure or purpose. (e.g, Differentiating, Organizing)&According to the graph, how does the colonisation pattern of the French Empire compare to that of the British Empire? Select all that apply. & The British Empire always had more colonies than the French Empire; The British Empire did not lose colonies before 1900, while the French Empire did; \textbf{The British and French Empires started to decline simultaneously}; The British and French Empires reach their peak number of colonies in the same year; The British Empire gained overseas colonies years before the French Empire \\ 
\midrule
\textit{Evaluate}: Making judgments based on criteria and standards. (e.g., Critiquing) & What other pieces of information would you need to evaluate the validity of the following sentence: ``\textit{The Spanish Empire impacted negatively more people than the French Empire}''. Select only the relevant options. & The duration of colonialism under the French and Spanish Empires; The number of colonies over time for the British and French Empires; \textbf{The population size of the colonies under the French and Spanish Empires}; \textbf{Economic and cultural policies implemented by the French and Spanish Empires in their colonies}; The information on the graph is sufficient \\  \midrule
\textit{Create}: Putting elements together to form a novel, coherent whole or make an original product. (e.g., Generating, Planning, Producing) & Based on the visualisations, write a text on the relationship between the topics presented in the graphs. Your response should be 8 to 12 sentences long and should reference specific data points or trends from the graphs to support your ideas. & \textbf{[Open Ended Task]} Example of an extract of an answer:      \textit{The two figures can be linked by the historical trends.  Figure 1 shaped the data shown in Figure 2. For example, Figure 1 shows the rise and fall of Spanish colonialism. We see likely downstream consequences of this reflected in Figure 2, with certain countries having higher rates of uncontrolled territory, as with Mexico, Peru, Colombia, Venezuela.} \\
\bottomrule
\end{tabular}
\label{tab:question_table}
\end{table}

\newpage
\section{Create Task Labelling Rubric}
\label{sec:appendix-create-rubric}

\begin{table}[h]
\centering
\caption{Rubric used to categorise sentences from participants' responses in the \textit{Create} task, based on the cognitive levels of Bloom's Taxonomy. }
\small
\begin{tabular}{p{1.5cm}p{5cm}p{8cm}}
\toprule
\textbf{Level} & \textbf{Description} & \textbf{Example} \\ \midrule
\textit{Identify} & Factual information from a graph without interpretation. Reference to values and information in one graph. & For example, the Spanish started colonizing the Americas in the 1400s and the British and French in the 1600s as can be seen from the first graph. \\ \midrule
\textit{Understand} & Explaining only one graph and interpreting the values. & The factors that affect GDP of a country including civil wars, politics, government policies, resources, investments and local and international trade. \\ \midrule
\textit{Apply} & Applying information from a graph to a new situation or hypothetical scenario about future trends. & For Syria, the civil war is still ongoing, so it is likely their state capacity is still low in 2022. \\ \midrule
\textit{Analysis} & Explanation linking both topics (reference to two graphs). Identifying potential relationships between the two topics. & The length of time that former colonies have been independent correlates with how much control those countries have over their territory. \\ \midrule
\textit{Evaluate} & Critical assessment of information and underlying assumptions in the graphs. Highlighting discrepancies, outliers or missing elements that affect interpretation. & More information is needed to validate the relationship as the first graph provides information until 2010 and the second graph provides information of 2022. \\ \midrule
\textit{Create} & Extending or extrapolating beyond the information provided in the graph to draw new conclusions, or propose alternative interpretations in the form of a new argument. & At the same time, rather than stability, countries with high rates of controlled land (that were post-colonial), such as South Africa, might not necessarily be a positive thing as colonialism imposed cultural, social, political norms, perhaps they had less ability for self-determination (versus other countries that had more ability to do so, even if it means higher rates of uncontrolled territory). \\ \midrule
\textit{Other} & Information not directly related to the graph to introduce another sentence (for example: topic sentences). Sentences of previous knowledge. & It's interesting to compare these visualisations between these two countries. \\ \bottomrule
\end{tabular}
\label{tab:levels}
\end{table}

\end{document}